\documentclass[]{aa}  

\usepackage{lscape}

\usepackage{graphicx}
\usepackage{txfonts}
\usepackage{acronym}
\acrodef{PC}[PC]{phase curve}
\acrodef{LC}[LC]{light curve}
\acrodef{LD}[LD]{limb darkening}
\acrodef{GP}[GP]{Gaussian Process}
\acrodef{MCMC}[MCMC]{Markov Chain Monte Carlo}
\acrodef{DRP}[DRP]{Data Reduction Pipeline}
\acrodef{PSF}[PSF]{Point Spread Function}
\acrodef{TTV}[TTV]{Transit Time Variation}
\acrodef{SHO}[SHO]{Simple Harmonic Oscillator}
\acrodef{BIC}[BIC]{Bayesian Information Criterion}
\acrodef{PSD}[PSD]{Power Spectral Density}
\acrodef{LS}[LS]{Lomb-Scargle}
\acrodef{ESP}[ESP]{Exponential-sine periodic}

\begin{document} 

    \graphicspath{{images/}}

   \title{Photometric follow-up of the 20 Myr-old multi-planet host star V1298~Tau with CHEOPS and ground-based telescopes\thanks{Based on data collected within the CHEOPS AO-3 proposal PR230032 \textit{Pinning down orbital period, transit ephemeris, and radius of the infant planet V1298\,Tau\,$e$. }}}


   \author{M.\,Damasso \inst{1}
          \and G.\,Scandariato\inst{2}
          \and V.\,Nascimbeni \inst{3}
          \and D.\,Nardiello \inst{3,4}
          \and L.\,Mancini \inst{5,1,6}
          \and G.\,Marino \inst{7}
          \and G.\,Bruno \inst{2}
          \and A.\,Brandeker \inst{8}
          \and G.\,Leto \inst{2}
          \and F.\,Marzari \inst{3}
          \and A.\,F.\,Lanza \inst{2}
          \and S.~Benatti \inst{9}
          \and S.~Desidera \inst{3}
          \and V.\,J.\,S.\,B\'ejar \inst{10,11}
          \and A.~Biagini \inst{12,9}
          \and L.~Borsato \inst{3}
          \and L.~Cabona \inst{3}
          \and R.~Claudi \inst{3,13}
          \and N.~Lodieu \inst{10,11}
          \and A.~Maggio \inst{9}
          \and M.~Mallorqu\'{i}n D\'{i}az \inst{10,11}
          \and S.~Messina \inst{2}
          \and G.~Micela \inst{9}
          \and D.~Ricci \inst{3}
          \and A.~Sozzetti \inst{1}
          \and A.\,Su\'{a}rez Mascare\~{n}o \inst{10,11}
          \and D.~Turrini \inst{1}
          \and M.\,R.\,Zapatero Osorio \inst{14}
          }
          

   \institute{INAF - Osservatorio Astrofisico di Torino, Via Osservatorio 20, I-10025 Pino Torinese, Italy\\
              \email{mario.damasso@inaf.it}
        \and INAF -- Osservatorio Astrofisico di Catania, Via S.~Sofia,78 - 95123 Catania, Italy
        \and  INAF -- Osservatorio Astronomico di Padova, Vicolo dell'Osservatorio 5, I-35122, Padova, Italy
        \and Aix Marseille Univ, CNRS, CNES, LAM, Marseille, France
        \and Department of Physics, University of Rome ``Tor Vergata'', Via della Ricerca Scientifica 1, 00133, Rome, Italy       
        \and Max Planck Institute for Astronomy, K\"{o}nigstuhl 17, 69117 -- Heidelberg, Germany   
        \and Gruppo Astrofili Catanesi, Catania, Italy
        \and Stockholm University, Department of Astronomy AlbaNova University Center, 106 91 Stockholm, Sweden
        \and INAF -- Osservatorio Astronomico di Palermo, Piazza del Parlamento 1, I-90134, Palermo, Italy
        \and Instituto de Astrof\'isica de Canarias (IAC), Calle V\'ia L\'actea s/n, E-38200 La Laguna, Tenerife, Spain 
        \and Departamento de Astrof\'isica, Universidad de La Laguna (ULL), E-38206 La Laguna, Tenerife, Spain
        \and Dipartimento di Fisica e Chimica E. Segré , Università di Palermo, Via Archirafi 36, 90100 Palermo, Italy
        \and Dipartimento di Matematica e Fisica, Universit\`a Roma Tre, Via della Vasca Navale 84, 00146 Roma, Italy
        \and Centro de Astrobiolog\'{\i}a CSIC-INTA, Carretera de Ajalvir km 4, E-28850 Torrej\'on de Ardoz, Madrid, Spain
             }

   \date{Received ; accepted }

 
  \abstract
   {The 20 Myr-old star V1298\,Tau hosts at least four planets. Since its discovery, this system has been a target of intensive photometric and spectroscopic monitoring. The characterisation of its architecture and planets' fundamental properties turned out to be very challenging so far.} 
   {The determination of the orbital ephemeris of the outermost planet V1298\,Tau\,$e$ remains an open question. Only two transits have been detected so far by $Kepler/K2$ and TESS, allowing for a grid of reference periods to be tested with new observations, without excluding the possibility of transit timing variations. Observing a third transit would allow to better constrain the orbital period, and would also help determining an accurate radius of V1298\,Tau\,$e$ because the former transits showed different depths.}
   {We observed V1298\,Tau with the CHaracterising ExOPlanet Satellite (CHEOPS) to search for a third transit of planet $e$ within observing windows that have been selected in order to test three of the shortest predicted orbital periods. We also collected ground-based observations to verify the result found with CHEOPS. We reanalysed $Kepler/K2$ and TESS light curves to test how the results derived from these data are affected by alternative photometric extraction and detrending methods.}
   {We report the detection with CHEOPS of a transit that could be attributed to V1298\,Tau\,$e$. If so, that result implies that the orbital period calculated from fitting a linear ephemeris to the three available transits is close to $\sim45$ days. Results from the ground-based follow-up marginally support this possibility. We found that \textit{i}) the transit observed by CHEOPS has a longer duration compared to that of the transits observed by $Kepler/K2$ and TESS; \textit{ii}) the transit observed by TESS is $>30\%$ deeper than that of $Kepler/K2$ and CHEOPS, and deeper than the measurement previously reported in the literature, according to our reanalysis.}
   {If the new transit detected by CHEOPS is due to V1298\,Tau\,$e$, this implies that the planet experiences TTVs of a few hours --as it can be deduced from three transits-- and orbital precession, which would explain the longer duration of the transit compared to the $Kepler/K2$ and TESS signals. Another, a priori less likely possibility is that the newly detected transit belongs to a fifth planet with a longer orbital period than that of V1298\,Tau\,$e$. Planning further photometric follow-up to search for additional transits is indeed necessary to solve the conundrum, also for pinning down the radius of V1298\,Tau\,$e$. }


    \titlerunning{Photometric follow-up of V1298\,Tau with CHEOPS and ground-based telescopes}
    \authorrunning{M.\,Damasso et al.}
   \maketitle
  
%
\section{Introduction} \label{sec:intro}
V1298\,Tau (also known as K2-309) is a $20\pm10$ Myr-old K1 star bolometric luminosity $L=0.954\pm0.040$ L$_{\odot}$; magnitude $V=10.12\pm0.05$; distance 109.5$\pm$0.7 pc; \citealt{Suarez_2022NatAs...6..232S}) that hosts four transiting planets discovered by \textit{Kepler/K2} \citep{david2019AJ....158...79D,David_2019ApJ...885L..12D}. This very young multi-planet system offers the unique opportunity of characterising planets at a very early phase after their formation. In turn, this allows us to constrain evolutionary models, whose theoretical predictions remain untested to date due to the paucity of such systems detected so far. Right after the discovery, V1298\,Tau became a target of intense multi-instrument, multi-band spectroscopic and photometric follow-up \citep{poppe2021MNRAS.500.4560P,feinstein2021AJ....162..213F,Suarez_2022NatAs...6..232S,vissapragada2021AJ....162..222V,gaidos2022MNRAS.509.2969G,marshall2022AJ....163..247J}. The study by \cite{Suarez_2022NatAs...6..232S} had the primary aim of measuring masses and bulk densities of the four planets in the V1298\,Tau system through spectroscopic observations. Their results show that V1298\,Tau\,b has a mass of $0.64\pm0.19$ $M_{\rm Jup}$ and a density similar to that of giant planets in the Solar System and other known old giant exoplanets. For the outermost V1298\,Tau\,$e$, for which only one transit was detected at that time, \cite{Suarez_2022NatAs...6..232S} determined an orbital period $P_e=40.2\pm0.1$ days, a mass of $1.16\pm0.30$ $M_{\rm Jup}$, and a density slightly larger, but compatible within error bars, than that of most of the older giant exoplanets. This unexpected result suggests that giant planets of such a young age might evolve and contract within the first million years after system's birth, thus challenging current models of planetary formation and evolution. Following the results of \cite{Suarez_2022NatAs...6..232S}, \cite{Maggio_2022ApJ...925..172M} investigated the current escape rates from the planetary atmospheres, and predicted the future evolution of atmospheric photo-evaporation, while \cite{Tejada_2022ApJ...932L..12T} investigated stability constraints on the system's orbital architecture.%

Despite a first and demanding follow-up, the results by \cite{Suarez_2022NatAs...6..232S} and their implications need to be confirmed, and an accurate characterisation of the V1298\,Tau system remains an open issue and a hot topic. In September-October 2021 the Transiting Exoplanet Survey Satellite (TESS) observed V1298\,Tau and detected a second transit attributed to planet $e$, as reported by \cite{Feinstein_2022ApJ...925L...2F}. They constrained the orbital period of the outermost planet to a grid of discrete values ($P_e>44$ days, not in agreement with the solution found by \citealt{Suarez_2022NatAs...6..232S} through spectroscopy), which is the only reasonable calculation when just two transits are detected. The discrete periods in the grid have been determined by dividing for integer numbers the time separation between the epochs of central transit observed by \textit{Kepler/K2} and TESS, and have to be interpreted as a guidance to search for future transits of planet $e$. They would correspond to actual values of $P_e$ if there are no substantial changes in the orbital parameters over short time scales, and without the presence of large transit timing variations (TTVs). The information currently available about the system does not allow us to reliably predict any significant orbital instability, and nowadays a transit follow-up strategy has to be necessarily based on the grid of orbital periods defined by \cite{Feinstein_2022ApJ...925L...2F}. 

\cite{Feinstein_2022ApJ...925L...2F} also showed that the transit depths of V1298\,Tau\,$bcd$ measured from TESS data are slightly shallower, but compatible within 1--2$\sigma$, than those observed by $Kepler/K2$, while the transit depth of V1298\,Tau\,$e$ is $\sim3\sigma$ larger. They conclude by proposing a few scenarios to qualitatively explain this discrepancy in radius measurements between $Kepler/K2$ and TESS, all of them involving phenomena not yet well characterised for systems with very young ages which make any interpretation very complex, going from the rapidly evolving stellar activity to haze-dominated planetary atmospheres. 
\cite{sikora2023} presented a reanalysis of the spectroscopic and photometric data of V1298\,Tau, including additional RVs to the dataset of \cite{Suarez_2022NatAs...6..232S}. Concerning V1298\,Tau\,$e$, and with a reference to the grid of guidance orbital periods derived by \cite{Feinstein_2022ApJ...925L...2F}, \cite{sikora2023} conclude that periods $P_e>55.4$ days can be ruled out at 3$\sigma$ level, and placed a tight constraint to $P_e$ ($P_e$ = 46.768131$\pm$0.000076 days assuming a circular orbit, although the actual orbital architecture of V1298\,Tau\,$e$ remains puzzling. 

In order to improve the characterisation of this system, we followed-up V1298\,Tau with the CHaracterising ExOPlanet Satellite (CHEOPS) space telescope \citep{Benz2021ExA....51..109B}. Our aim was twofold\footnote{We based our proposal on the hypothesis by \cite{Feinstein_2022ApJ...925L...2F} that the two individual transits observed by $Kepler/K2$ and TESS belong to the same planetary companion V1298\,Tau\,$e$}: \textit{i}) detecting additional transits of planet $e$, providing better constraints to its orbital period and a first identification of TTVs, with the consequent improvement of the results by \cite{Feinstein_2022ApJ...925L...2F}, and \textit{ii}) getting a new measurement of its radius, possibly helping to explain the different transit depths measured from $Kepler/K2$ and TESS \acp{LC}. In this work, we present results from the CHEOPS monitoring conducted between 18 November and 18 December 2022, and from further ground-based photometric follow-up that was motivated by the outcomes of CHEOPS observations. We also reanalysed the TESS photometry that we extracted using an alternative pipeline to correct for contamination from nearby faint
stars, and compare the results to those of \cite{Feinstein_2022ApJ...925L...2F}.   
\section{Description of the photometric data} \label{sec:data}
\subsection{Kepler/K2}
We used the \texttt{EVEREST 2.0} \citep{2018AJ....156...99L} version of the $Kepler/K2$ LC, also analysed by \cite{David_2019ApJ...885L..12D} and \cite{Feinstein_2022ApJ...925L...2F}. The time series covers a time span of about 71 d from 8 February 2015 to 20 April 2015 ($Kepler/K2$ Campaign 4).
\subsection{TESS} \label{sec:datatess}
V1298\,Tau was observed by TESS between 16 September and 6 November 2021 (Sectors 43 and 44), and the observations were analysed shortly after by \cite{Feinstein_2022ApJ...925L...2F}. We re-extracted the LC from the TESS Full Frame Images (FFIs) with the PATHOS pipeline described by \citet{2019MNRAS.490.3806N,2020MNRAS.495.4924N}. First, we extracted the long-cadence LC (cadence of 10 minutes) by using six different photometric apertures (PSF-fitting, 1-px, 2-px, 3-px, 4-px, and 5-px radius aperture photometry) after subtracting from each FFI the sources within 3~arcmin from V1298\,Tau. Then, we corrected the LC by applying the cotrending basis vectors (CBVs) calculated as in \citet{2021MNRAS.505.3767N}. We selected the best photometric aperture comparing the point-to-point (P2P) rms of the different \acp{LC}, finding that the lower P2P rms is that associated with the PSF-fitting light curve. V1298~Tau is an ideal target to be analysed with the PATHOS pipeline, that was developed to extract high-precision light curves in crowded fields, like open clusters. The PSF-fitting photometry performed with empirical PSFs allow us to minimise the dilution effects due to the contamination of close-by stars (following the TIC catalogue, there are 63 sources within 3 arcmin, two of them having $T \sim 9.6$ and $T\sim 7.9$). Moreover, in \cite{2022A&A...664A.163N}, we discussed that our CBV correction of systematic errors in the light curves is expected to perform more effectively than other pipelines in the case of active stars such as V1298~Tau (no spurious signals or transit deformations occurred in our final light curves).

\subsection{CHEOPS} \label{sec:cheopsobs}
\begin{figure}
    \centering
    \includegraphics[trim={50 470 0 100},clip,width=0.55\textwidth]{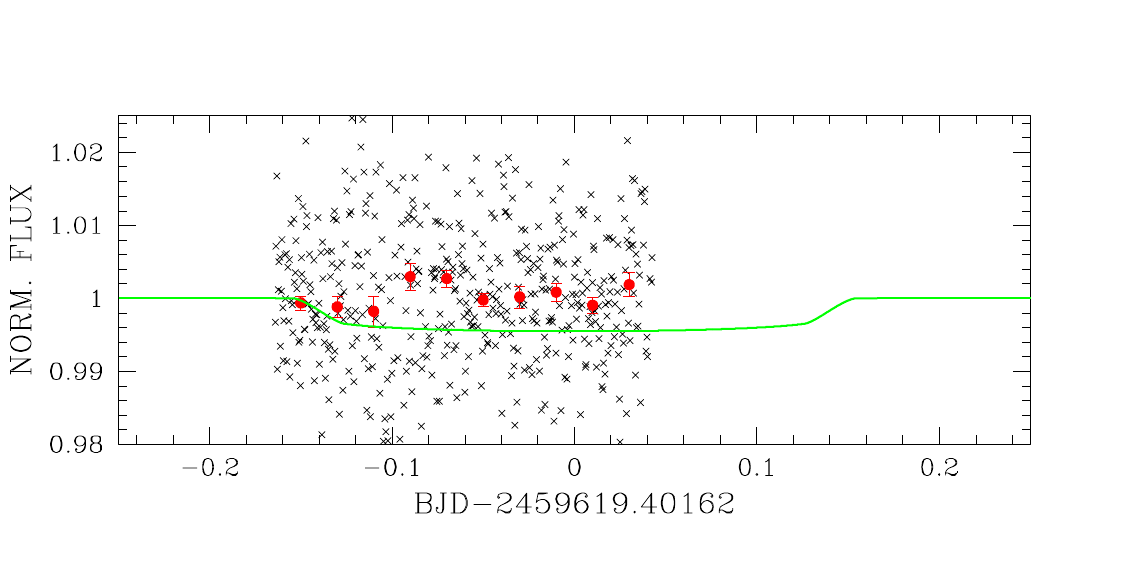}
    \caption{Light curve of V1298\,Tau collected on the night of 8 February 2022 with the Schmidt 67/92cm telescope at the INAF-Astrophysical Observatory of Asiago. The red dots corresponds to the binned light curve, and the green curve represents the transit model of planet $e$ based on the ephemeris and transit depth derived by \cite{Feinstein_2022ApJ...925L...2F}.}
    \label{fig:asiago8febbraio2022}
\end{figure}

We were awarded a total of 112 CHEOPS orbits to test three orbital periods of V1298\,Tau\,$e$ with the highest probability among those calculated by \cite{Feinstein_2022ApJ...925L...2F} from the temporal separation between the two transits seen by \textit{Kepler/K2} and TESS, namely 44.1699, 45.0033, and 46.7681 ($\pm0.0001$) days. It must be emphasised that the grid of periods proposed by \cite{Feinstein_2022ApJ...925L...2F} originates from just two observed transits, and they can be used as a guidance to search for a third transit that would better constraint the orbital ephemeris. Our original plan was dividing the 112 orbits in six visits, two visits per test period, in order to collect two transits in the case of a successful detection, and get a robust measurement of the orbital period and planet's radius. In practice, three over six visits have been executed. We did not select $P$=45.8687 days as a test period from the grid because only one CHEOPS visit could be accommodated in the available time span, due to observing constraints. Concerning this period, we did not find clear evidence for a transit event in photometry collected between BJD$_{\rm TDB}$ 2\,459\,619.2380 and 2\,459\,619.4459 (8 February 2022) with the Schmidt 67/92cm telescope at the INAF-Astrophysical Observatory of Asiago, when a transit was predicted with ingress, centre, and egress BJD$_{\rm TDB}$ 2\,459\,619.24662, 2\,459\,619.40162 and 2\,459\,619.55662 respectively. This LC is shown in Fig. \ref{fig:asiago8febbraio2022} and was extracted as described in \cite{nardiello2015MNRAS.447.3536N,nardiello2016MNRAS.455.2337N}. Looking at the data, we cannot claim a detection with a significance better than nearly 2$\sigma$. Also, possible large TTVs should be taken into account before discarding $P$=45.8687 days from the list of the possible guidance orbital periods. The TTVs cannot be reliably predicted yet because, as already mentioned, many critical information about the system are still missing. Nonetheless, we found convenient for the sake of our proposal to focus on the other three high-probability test periods that we selected. The observing windows have been centred to the predicted time of central transit assuming no TTVs, and their time span (nearly 1 or 2 days) has been defined to allocate to some extent possible, but still unknown TTVs. 

We extracted the LC using the \ac{PSF} photometry package PIPE\footnote{\url{https://github.com/alphapsa/PIPE}} \citep{Brandeker2022}, which is suitable for crowded fields. This pipeline models the background stars before extracting the photometry of the target by fitting a \ac{PSF}. The \ac{PSF} fitting is weighted by the signal and noise of each pixel, thus the extraction is less sensitive than aperture photometry to background star contamination. We use this curve in the present analysis. A comparison with the default \ac{DRP} of CHEOPS \citep{hoyer2020A&A...635A..24H} reveals a general 30\% improvement in the overall photometric scatter. However, as a cross-check we also analysed the \ac{DRP} photometry, finding no differences in the results within the uncertainties.\\

\begin{figure}
    \centering
    \includegraphics[trim={0 300 0 200},clip, width=0.5\textwidth]{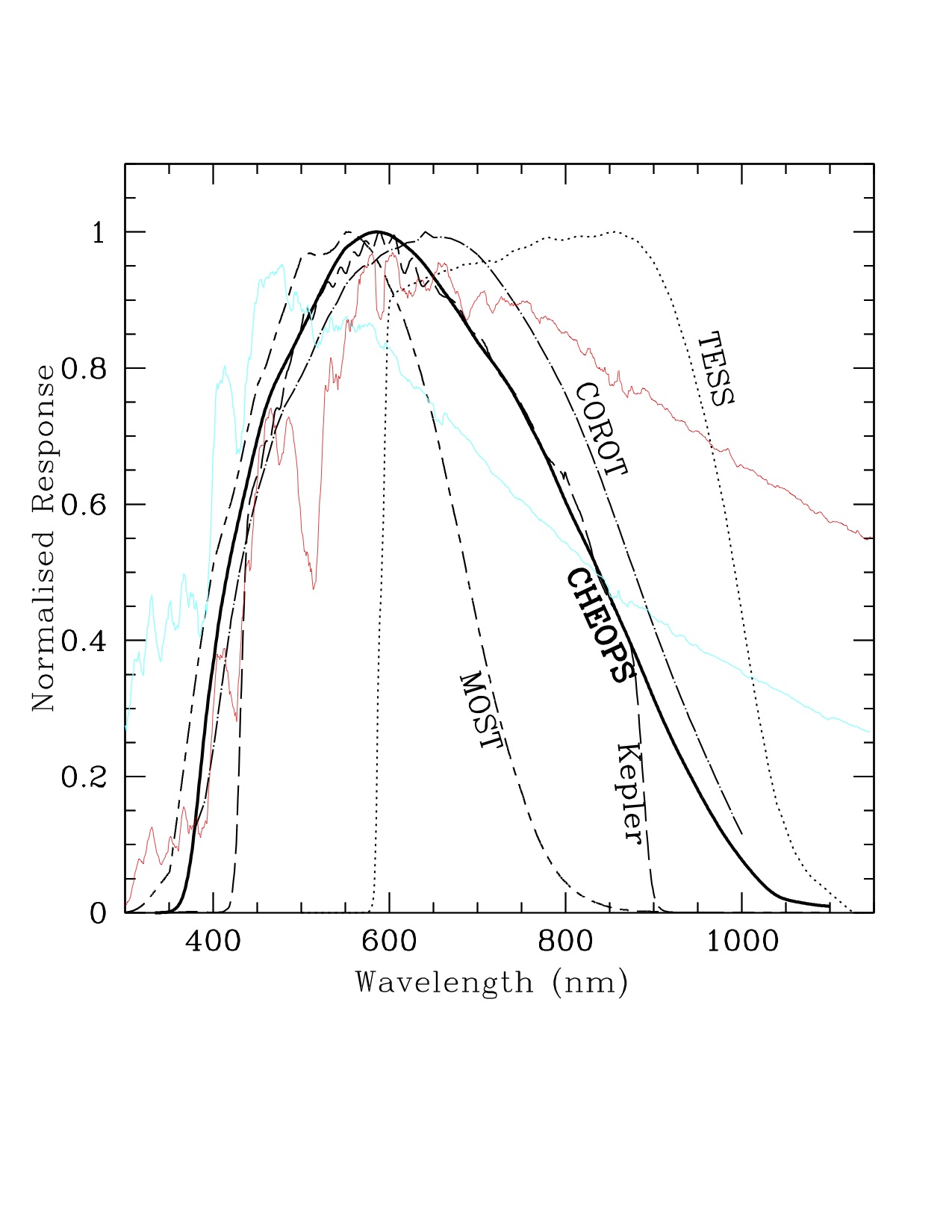}
    \caption{Bandpasses of $Kepler/K2$, TESS, and CHEOPS normalised to their peak values, as a function of wavelength. The cyan and red curves represent the spectral energy distributions of a T$_{\rm eff}$=5500 K and T$_{\rm eff}$=4500 K dwarf star, respectively (taken from https://www.cosmos.esa.int/web/cheops/performances-bandpass).}
    \label{fig:passbands}
\end{figure}

For convenience, we show the bandpasses of the three space telescopes in Fig. \ref{fig:passbands}.

\subsection{Ground-based observations}\label{sec:ground}
V1298\,Tau was observed during the night of 31 January - 01 February 2023 by ground-based facilities in Italy and Spain. This joint follow-up was aimed to support the findings of CHEOPS described in Sect. \ref{sec:lcmodelling}, and the involved facilities are described in the following:
\begin{itemize}
\item \emph{Asiago Copernico.} We gathered a photometric series through a Sloan $r$ filter with the AFOSC instrument (Asiago Faint Object Spectrograph and Camera) mounted at the 1.82-m Copernico telescope at Cima Ekar, in northern Italy\footnote{\url{https://www.oapd.inaf.it/sede-di-asiago/telescopes-and-instrumentations/copernico-182cm-telescope}}. A total of 3\,364 frames were collected with a constant exposure time of 5~s from 17:07~UT to 00:18~UT, when the target elevation went below the safety limit. The sky was mostly clear throughout the series with a few passing thin veils. The 81\%-illuminated Moon at just $16^\circ$ from the line of sight resulted in a very high background level and a scatter larger than usual, but with no significant impact on photometric accuracy. The images, purposely defocused to $\sim 7''$~FWHM to improve photometric accuracy and avoid saturation, were bias- and flat-field corrected using standard procedures. A LC was then extracted with the STARSKY code \citep{Nascimbeni2013}, a photometric pipeline developed by the TASTE project \citep{Nascimbeni2011} to perform high-precision differential aperture photometry over defocused images. 
\item \emph{Asiago Schmidt.} We also acquired a photometric series through a Sloan $i$ filter with a second telescope at the same observing site, the Asiago 67/92~cm Schmidt camera operating in robotic mode\footnote{\url{https://www.oapd.inaf.it/sede-di-asiago/telescopes-and-instrumentations/schmidt-6792}}. A total of 836 frames were collected with a constant exposure time of 7~s from 17:28~UT to 00:04~UT, with two $\sim 30$~min gaps in the first half due to technical issues. A differential LC was extracted using the same STARSKY procedure applied to the Asiago Copernico data.
\item \emph{Calar Alto.} V1298\,Tau was also observed with the Zeiss 1.23\,m telescope at the observatory of Calar Alto (Spain), using a Cousins-$I$ filter. We defocussed the telescope to improve the quality of the photometric data, a technique often used for transit follow-up \citep{2009MNRAS.396.1023S}. We used unbinned and windowed CCD (DLR-MKIII CCD camera\footnote{http://www.caha.es/telescope-1-23m/ccd-camera}, field of view $21.4^{\prime} \times 21.4^{\prime}$) in order to decrease the saturation threshold and the readout time. The monitoring started on 31 January 2023 18:45 UT (airmass 1.06), and ended on 01 February 2023 00:24 UT (airmass 2.31). The exposure time was gradually increased from 30 to 75 sec with the increase of the airmass. The images were reduced using master-flat and master bias frames which were collected before the scientific observations. The calibrated time series were corrected for pointing variations, and aperture differential photometry was performed with the \texttt{APER} routine, which is part of the \texttt{IDL} \texttt{ASTROLIB} library \citep{astrolib1993ASPC...52..246L}.
\item \emph{Catania.} Photometric observations were collected with the 25-cm Newtonian telescope of Gruppo Astrofili Catanesi (Catania, Italy; lon.~15:05:32 E, lat.~+37:30:23, alt.~42~m) using a Johnson-Cousins $R$ filter. The telescope was equipped with a coma corrector and a CCD camera Sbig ST-7 XME. Overall, 243 frames were acquired from 17:37~UT to 00:33~UT with a constant exposure time of 100~s, chosen to reduce the
effects of scintillation. To optimise the SNR, a slight defocus was used, obtaining a FWHM of about 7''.
After standard calibration of each image with bias, dark and flat field frames, the light curve was obtained via not-weighted ensemble aperture photometry with the software \texttt{AstroImage J} \citep{2017AJ....153...77C}.
\end{itemize}


\section{Light curve modelling} \label{sec:lcmodelling}

We analysed the \acp{LC} of all instruments in a Bayesian framework. 
We used the \texttt{python} package \texttt{SnappyKO} \citep{Parviainen2020} to model the transits. Correlated signals due to stellar activity are modelled through a \ac{GP} regression using the \texttt{S+LEAF}  \texttt{python} library \citep{Delisle2020,Delisle2022}. All the details of the GP modelling are provided in Sect. \ref{sec:activity}.

The likelihood maximisation is performed with a \ac{MCMC} using the package \texttt{emcee} version 3.1.3 \citep{Foreman2013}, a \texttt{python} implementation of the affine invariant MCMC ensemble sampler by \cite{Goodman2010CAMCS...5...65G}. We ran the code in the HOTCAT computing infrastructure \citep{Bertocco2020,Taffoni2020}, to speed-up the computational time.

\subsection{$Kepler/K2$ light curve}\label{sec:k2Analysis}

We masked the transits of planets $bcd$ from the $Kepler/K2$ LC by computing the time of transits using the ephemeris of \citet{David_2019ApJ...885L..12D} and trimming a time window of 1.5 times the corresponding transit duration centred on the expected transit time. Then, we flattened the LC using the best-fit model of \citet{David_2019ApJ...885L..12D}, in order to identify and reject stellar flares (using a +5$\sigma$ clipping threshold above the best-fit model).


As discussed in Appendix \ref{sec:activity}, we found that the best GP model is a combination of a quasi-periodic Exponential-sine periodic (ESP) kernel to trace the rotational signal, and stochastically-driven harmonic oscillator (SHO) kernel to modelling the low-frequency correlated noise, which is likely of stellar origin, as it is recovered also in the analysis of the TESS \ac{LC} (see Sect.~\ref{sec:tessAnalysis}).

We fitted the data using a model containing simultaneously the transit signal of planet $e$ (assuming a circular orbit) and the correlated stellar rotation signal. We used the priors listed in Table~\ref{tab:k2TessFit}. For the orbital period $P_e$ we used a uniform prior $\mathcal{U}$(35, 100) days, and for the time of central transit we used a uniform prior $\mathcal{U}(2\,457\,096.46, 2\,457\,096.79)$ BJD$_{\rm TDB}$ centred on the epoch of transit indicated by \citet{David_2019ApJ...885L..12D}. We used the stellar density instead of fitting directly the scaled semi-major axis $a/R_*$, adopting a Gaussian prior based on the measurement of \citet{Suarez_2022NatAs...6..232S}. To fit the \ac{LD} coefficients (adopting a quadratic law), we used priors based on the results of \citet{David_2019ApJ...885L..12D}. The $Kepler/K2$ \ac{LC} has a cadence of 29.4 minutes. Following a common practice, in order to mitigate morphological distortions introduced when fitting transit light curves with such a long-cadence (see, e.g., \citealt{kipping2010}), we used an oversampling factor of 3. This means that each simulated photometric data point at the epoch of a real observation is calculated as the average value of three evenly spaced simulated data points each corresponding to a 10-minute exposure.

We ran the \ac{MCMC} for 100\,000 steps, which turned out to be $\sim$150 times the auto-correlation length of the chains, estimated following \citet{Goodman2010CAMCS...5...65G}. This indicated successful convergence, as suggested by \citet{Sokal1997} and adapted to parallel Monte Carlo chains (see \url{https://dfm.io/posts/autocorr/}). The best-fit model of the $Kepler/K2$ \ac{LC} is shown in Fig.~\ref{fig:LCs}, while in Fig.~\ref{fig:transits} we show the detrended transit signal (black dots). We obtained a transit best-fit solution consistent with \citet{David_2019ApJ...885L..12D} within $1\sigma$ (Table~\ref{tab:k2TessFit}).

\subsection{TESS light curve}\label{sec:tessAnalysis}

For the fit of the TESS \ac{LC} we used the same approach adopted for the $Kepler/K2$ data, as described above. Also in this case, we found out that a mixture of a ESP and SHO kernels provides a more realistic representation of the correlated noise in the \ac{LC}, and through several tests we found that using different kernels does not affect the transit parameters. Following the results of \citet{Feinstein_2022ApJ...925L...2F}, we used the prior $\mathcal{U}(2459481.63, 2459481.96)$ BJD$_{\rm TDB}$ for the time of transit, and a Gaussian prior for the \ac{LD} coefficients. We used an oversampling factor of 3 to model the data with a cadence of 10 min.
The best-fit model and the detrended transit signal are shown in Fig.~\ref{fig:LCs} and Fig.~\ref{fig:transits} (orange dots), respectively. For comparison, we also show in Fig. \ref{fig:transits} the TESS transit measured by \cite{Feinstein_2022ApJ...925L...2F} (green data points), which has a lower depth. This discrepancy is due to both the different LC extraction methods and GP regression analysis, as it is evident looking at Fig. \ref{fig:tessgpcomparison}. The two LCs looks very similar, except for the out of transit segments, and our GP-modelled flux is higher during the transit timespan, overall determining a deeper transit in our case. 

\subsection{CHEOPS light curve} \label{sec:cheopslcanalysis}

We modelled the LC of all the CHEOPS visits using the same framework described above and taking advantage of some of the results from the previous analysis. First of all, we found that a simple undamped SHO kernel could effectively mitigate the correlated signal present in the \ac{LC} that is due to the stellar rotation. This is expected from the fact that the \ac{LS} periodogram of the data shows only the rotational peak, not its harmonics. Moreover, the adoption of the SHO kernel minimises the number of free parameters compared to the ESP kernel. We assumed that the low-frequency correlated noise detected in the $Kepler/K2$ and TESS data could not be recovered in CHEOPS data due to the limited time span and sparse sampling. Thus, we fitted the GP correlated noise using  a Gaussian prior for the stellar rotation frequency $\nu_{\rm rot}$ measured in the TESS data (corresponding to the rotation period of $\sim2.9$ days). For the \ac{LD} coefficients we used Gaussian priors based on the results of \citet{David_2019ApJ...885L..12D}, because Kepler and CHEOPS passbands are similar (Fig. \ref{fig:passbands}, and \citealt{Benz2021ExA....51..109B}). The prior on the time of inferior conjunction $T_0$ is uniform, and spans the full extent of the CHEOPS follow-up ($\mathcal{U}(2\,459\,901.8,2\,459\,932.5)$ BJD$_{\rm TDB}$) in order to keep the search for a transit signal blind. We adopted a uniform prior $\mathcal{U}$(35, 100) days for the orbital period of a possible transiting companion. 

The CHEOPS \acp{LC} are affected by periodic instrumental noise due to the roll of the telescope. We modelled this systematic effect jointly with the transit and stellar rotational signals by using a harmonic expansion up to the fifth harmonic of a periodic signal phased with the roll angle of the telescope, as described in \cite{Scandariato2022}. A posteriori, we also found some residual correlation with the coordinates ($x_{\rm PSF},y_{\rm PSF}$) of the centroid of the stellar \ac{PSF}, which we corrected by including in the model a bi-linear detrending in $x_{\rm PSF}$ and $y_{\rm PSF}$. 

\begin{figure*} 
    \centering
    \includegraphics[angle=0,width=\linewidth]{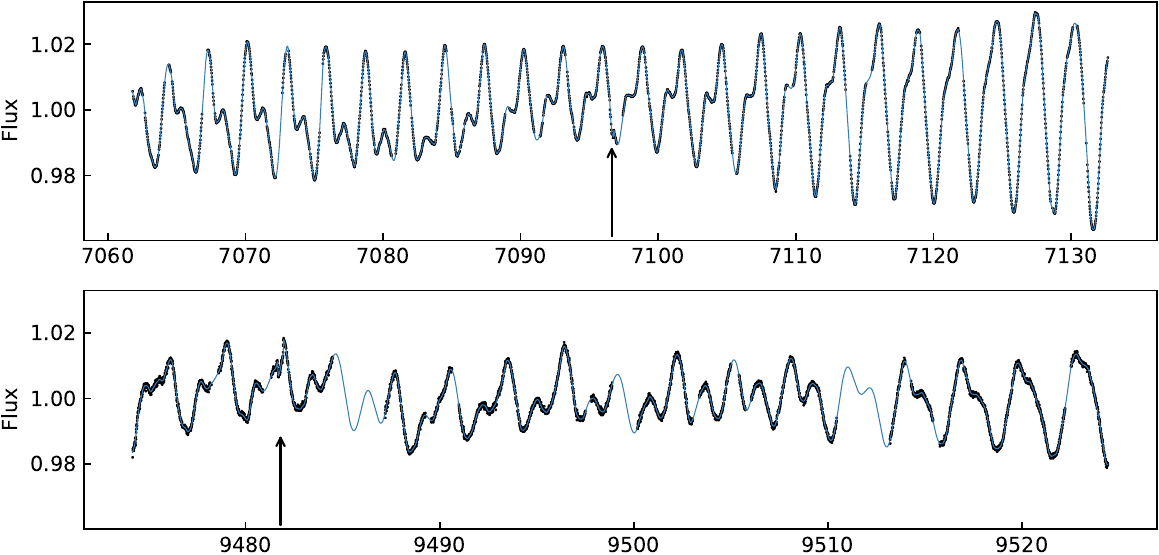}\\
    \includegraphics[angle=0,width=\linewidth]{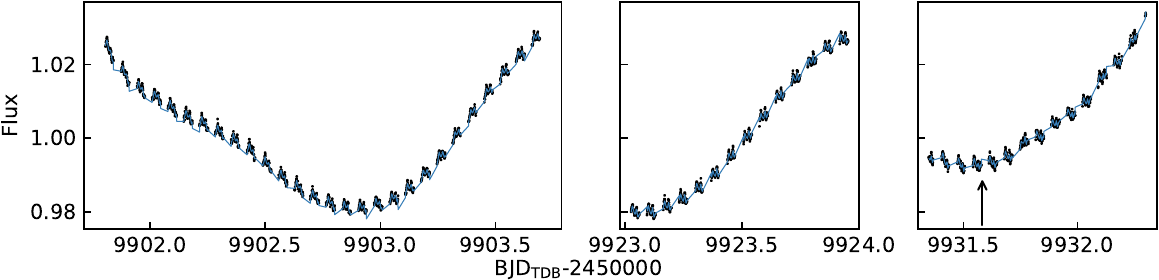}
    \caption{$Kepler/K2$, TESS and CHEOPS \acp{LC} of V\,1298\,Tau from top to bottom row respectively. In each panel, the blue solid line shows the best-fit model. The arrows indicate the transits of the companion labelled as planet $e$ by \cite{David_2019ApJ...885L..12D} and \cite{Feinstein_2022ApJ...925L...2F}, based on data from K2 and TESS.}
    \label{fig:LCs}
\end{figure*}

The results of this analysis are \textit{i}) the non-detection of transit-like signals in the first and second visit. This result puts some constraints on the orbit of the planet V1298\,Tau\,$e$, but it is not conclusive, and it does not allow us to exclude that the orbital period, assuming the presence of TTVs not yet verified, is close to $P$=44.1699 and 46.7681 days \citep{Feinstein_2022ApJ...925L...2F}. This point is further discussed in Section 4; \textit{ii}) the detection of a transit-like signal in the first half of the third CHEOPS slot (last panel in Fig. \ref{fig:LCs}), where a transit of V1298\,Tau\,$e$ is expected to occur if the orbital period were close to 45.0033 days, assuming some TTVs.
During the third visit, there were no predicted transits of any of the other three planets, well within the uncertainties, according to the ephemeris calculated by \cite{Feinstein_2022ApJ...925L...2F}. The \ac{BIC} of this model is -34064. Then, we fitted the data without including the transit signal. The corresponding solution has a BIC=-34048 ($\Delta$BIC=+16), thus this model is highly disfavoured with respect to the one that includes the transit.

\begin{figure*}
    \centering
    \includegraphics[width=\linewidth]{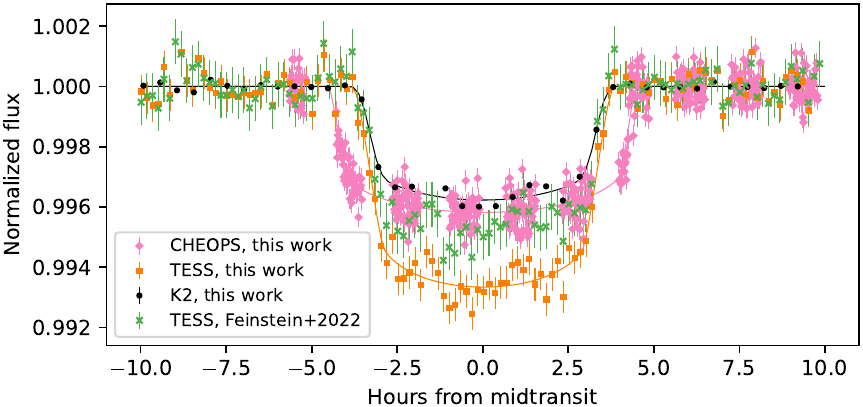}
    \caption{Comparison of the detrended transit \acp{LC} discussed in this work as seen by $Kepler/K2$, TESS and CHEOPS. For each dataset, the best-fit transit model is shown by curves with the same colour as the data points. The green data points identify the detrended and flattened TESS light curve analysed by \cite{Feinstein_2022ApJ...925L...2F}, obtained from the \texttt{jupyter} notebook made publicly available by the authors at https://github.com/afeinstein20/v1298tau$\_$tess/blob/main/notebooks/TESS$\_$V1298Tau.ipynb}
    \label{fig:transits}
\end{figure*}

We derive a transit duration $T_{14}\sim9$ hr and an impact parameter $b=0.11^{+0.09}_{-0.07}$ which are, respectively, longer and shorter than those measured from $Kepler/K2$ and TESS data (Table~\ref{tab:cheopsFit} and Fig. \ref{fig:transits}). Specifically, the discrepancies between the transit duration measured by CHEOPS and the duration observed by $Kepler/K2$ and TESS are ($T_{\rm 14,\,CHEOPS}$-$T_{\rm 14,\,K2}$)/$\sqrt{(\sigma_{\rm T_{14},\,CHEOPS}^2+\sigma_{\rm T_{14},\,K2}^2)}$ =$\Delta(T_{\rm 14,\,CHEOPS,\,K2})/\sqrt{\Sigma(\sigma_{\rm T_{14,\,CHEOPS,\,K2}}^2)}$=14.2, and $\Delta(T_{\rm 14,\,CHEOPS,\,TESS})/\sqrt{\Sigma(\sigma_{\rm T_{14,\,CHEOPS,\,TESS}}^2)}$=10.6, respectively (in case of values with asymmetric error bars, in the calculation we used the average values of the upper and lower uncertainties). For the impact parameter, the discrepancies between the measurements are $\Delta(b_{\rm CHEOPS,\,K2})/\sqrt{\Sigma(\sigma_{\rm b_{\rm CHEOPS,\,K2}}^2)}$=3.7, and $\Delta(b_{\rm CHEOPS,\,TESS})/\sqrt{\Sigma(\sigma_{\rm b_{\rm CHEOPS,\,TESS}}^2)}$=5.2 Our value of the orbital period $P_{\rm orb}=58\pm7$ days is consistent within 2$\sigma$ with the value of 45 days expected for the third visit by CHEOPS. The derived transit depth $R_p^2/R_\star^2= 0.0036\pm0.006$ is in very good agreement with that measured from $Kepler/K2$ photometry, $\Delta(R_{\rm p,\,CHEOPS,\,K2}^2/R_\star^2$)/$\sqrt{\Sigma(\sigma_{\rm R_{\rm p,\,CHEOPS,\,K2}^2/R_\star^2}^2)}$=0.1, but differs more significantly from that we derived from our re-processed TESS LC\footnote{\citealt{Feinstein_2022ApJ...925L...2F} found $R_p/R_\star=0.0664^{+0.0025}_{-0.0021}$ using their version of TESS photometry, corresponding to a transit depth 0.0043$\pm$0.0003}, $\Delta(R_{\rm p,\,CHEOPS,\,TESS}^2/R_\star^2$)/$\sqrt{\Sigma(\sigma_{\rm R_{\rm p,\,CHEOPS,\,TESS}^2/R_\star^2}^2)}$=4.5. The time of central transit occurs $\sim$6 hrs earlier than the epoch predicted by the ephemeris of \cite{Feinstein_2022ApJ...925L...2F} (BJD$_{\rm TDB}$ 2\,459\,931.8299$\pm$0.0015, corresponding to transit number 10 after the TESS observations). In principle, significant TTVs could not be ruled out, and would explain the significant discrepancy with the predicted epoch $T_0$. We note that the best-fit timescale of the SHO kernel is $\sim$2 days (Table~\ref{tab:cheopsFit}), meaning that the \ac{GP} model cannot filter out features of the \ac{LC} with shorter timescales, such as the planetary ingress and egress. Thus, we conclude that the \ac{GP} is not responsible for altering the shape of the transit signal. We show in Fig. \ref{fig:threetransitsgp} the original transit LCs gathered by the three telescopes that we analysed in this work, with overplotted our best-fit GP models. 

As a final test, we fit the CHEOPS \ac{LC} 
but substituting the \ac{GP} detrending with a fifth-degree polynomial (a polynomial for each CHEOPS visit). The polynomial detrending does not perform as effectively as the GP in this case, as it is confirmed by the presence of correlated noise in the residuals of the best-fit. This introduces biases in the estimated best-fit parameters, but the advantage is that polynomials are less prone to absorb astrophysical signals or introduce false positives than \acp{GP}. Following this alternative approach, we still found a transit signal in the first half of the third visit which is again significantly favoured by the BIC over the model without a transit. The best-fit parameters, as said above, are less reliable due to the worse quality of the detrending, nonetheless in both cases (fixed and free orbital period) the transit duration is longer than that observed by $Kepler/K2$ and TESS.

\begin{sidewaystable*}
\tiny
\caption{Model parameters for the fit of the $Kepler/K2$ and TESS data.\tablefootmark{a}}\label{tab:k2TessFit}
\renewcommand{\arraystretch}{1.5} 
\begin{tabular}{lll|ccc|ccc|l}
\hline\hline
Jump parameters & Symbol & Units & \multicolumn{3}{c|}{K2} & \multicolumn{3}{c|}{TESS} & Prior\\
                &        &       &  free $P_{\rm orb}$  &  $P_{\rm orb}=45.0033$ d & $P_{\rm orb}=45.0033$ d & free $P_{\rm orb}$  &  $P_{\rm orb}=45.0033$ d & $P_{\rm orb}=45.0033$ d\\
                &        &       &  circular orbit  &  circular orbit & eccentric orbit & circular orbit  &  circular orbit & eccentric orbit\\
\hline
\textit{ESP kernel}\\
Amplitude scale & $\log \sigma_{\rm ESP}$ &  & -4.0$^{+0.2}_{-0.2}$ & -4.0$^{+0.2}_{-0.2}$ & -4.0$^{+0.2}_{-0.2}$ & -4.5$^{+0.3}_{-0.2}$ & -4.5$^{+0.3}_{-0.3}$ & -3.9$^{+1.0}_{-0.8}$ & $\mathcal{U}$(-6,-2)\\
Scale of the oscillations & \textbf{$\log \eta_{\rm ESP}$} &  & -0.6$^{+0.2}_{-0.2}$ & -0.6$^{+0.2}_{-0.2}$ & -0.6$^{+0.2}_{-0.2}$ & -1.1$^{+0.5}_{-0.9}$ & -1.1$^{+0.5}_{-0.9}$ & -0.1$^{+0.6}_{-1.3}$ & $\mathcal{U}$(-3,1)\\
Rotational frequency & $\nu_{\rm rot}$ & day$^{-1}$ & 0.347$^{+0.001}_{-0.001}$ & 0.347$^{+0.001}_{-0.001}$ & 0.347$^{+0.001}_{-0.001}$ & 0.345$^{+0.001}_{-0.001}$ & 0.345$^{+0.001}_{-0.001}$ & 0.346$^{+0.002}_{-0.001}$ & $\mathcal{U}$(0.3,0.4)\\
Timescale & $\log \frac{l_{\rm ESP}}{1~day}$ &  & 2.86$^{+0.10}_{-0.09}$ & 2.86$^{+0.10}_{-0.09}$ & 2.86$^{+0.10}_{-0.09}$ & 3.4$^{+0.3}_{-0.2}$ & 3.4$^{+0.3}_{-0.2}$ & 2.7$^{+0.7}_{-0.4}$ & $\mathcal{U}$(0,5)\\
\textit{SHO kernel} \\
Amplitude scale & $\log \sigma_{\rm SHO}$ &  & -6.39$^{+0.04}_{-0.04}$ & -6.39$^{+0.05}_{-0.04}$ & -6.39$^{+0.05}_{-0.04}$ & -5.72$^{+0.11}_{-0.10}$ & -5.7$^{+0.1}_{-0.1}$ & -6.5$^{+0.8}_{-0.3}$ & $\mathcal{U}$(-8,-5)\\
Primary frequency & $\nu_0$ & day$^{-1}$ & 1.03$^{+0.03}_{-0.03}$ & 1.03$^{+0.03}_{-0.03}$ & 1.03$^{+0.03}_{-0.03}$ & 0.47$^{+0.05}_{-0.05}$ & 0.47$^{+0.05}_{-0.05}$ & 1.0$^{+0.2}_{-0.5}$ & $\mathcal{U}$(0.01,5)\\
Timescale & $\log \frac{l_{\rm SHO}}{1~day}$ &  & -1.03$^{+0.08}_{-0.08}$ & -1.03$^{+0.08}_{-0.08}$ & -1.03$^{+0.08}_{-0.08}$ & -1.4$^{+0.1}_{-0.1}$ & -1.4$^{+0.1}_{-0.1}$ & -1.2$^{+0.3}_{-0.2}$ & $\mathcal{U}$(-4,0)\\
\textit{Stellar and planetary parameters} \\
Stellar density & $\rho_\star$ & $\rho_\sun$  & 0.57$^{+0.06}_{-0.06}$ & 0.57$^{+0.07}_{-0.07}$ & 0.57$^{+0.07}_{-0.07}$ & 0.56$^{+0.07}_{-0.07}$ & 0.51$^{+0.07}_{-0.06}$ & 0.55$^{+0.08}_{-0.07}$ & $\mathcal{N}$(0.56,0.10)\\
Time of inferior conjunction - 2\,450\,000 & $T_0$ & BJD$_{\rm TDB}$ & 7096.6211$^{+0.0007}_{-0.0007}$ & 7096.6211$^{+0.0007}_{-0.0007}$ & 7096.6211$^{+0.0007}_{-0.0007}$ & 9481.797$^{+0.001}_{-0.001}$ & 9481.797$^{+0.001}_{-0.001}$ & 9481.797$^{+0.001}_{-0.001}$ & see text\\
Orbital period & P$_{\rm orb}$ & day & 46$^{+10}_{-6}$ & -  & -  & 58$^{+12}_{-9}$ & -  & -  & $\mathcal{U}$(35,100)\\
Planet-to-star radius ratio & $R_p/R_\star$ &  & 0.059$^{+0.003}_{-0.002}$ & 0.059$^{+0.002}_{-0.002}$ & 0.059$^{+0.002}_{-0.002}$ & 0.079$^{+0.003}_{-0.003}$ & 0.078$^{+0.002}_{-0.002}$ & 0.079$^{+0.003}_{-0.003}$  & $\mathcal{U}$(0,0.2)\\
Impact parameter & $b$ &  & 0.54$^{+0.08}_{-0.09}$ & 0.52$^{+0.05}_{-0.06}$ & 0.5$^{+0.1}_{-0.2}$ & 0.61$^{+0.05}_{-0.06}$ & 0.55$^{+0.06}_{-0.07}$ & 0.60$^{+0.06}_{-0.07}$  & $\mathcal{U}$(0,0.8)\\
Orbital eccentricity & $e$ &  & -$^{}_{}$ & -$^{}_{}$ & 0.1$^{+0.2}_{-0.1}$ & -$^{}_{}$ & -$^{}_{}$ & 0.15$^{+0.18}_{-0.10}$  & $\mathcal{U}$(0,1)\\
Argument of periastron\tablefootmark{c} & $\omega_{\star}$ & rad & -$^{}_{}$ & -$^{}_{}$ & 0$^{+2}_{-3}$ & -$^{}_{}$ & -$^{}_{}$ & -1$^{+1}_{-1}$  &  $\mathcal{U}$(0,2$\pi$)\\
Linear LD coefficient & u$_1$ &  & 0.4$^{+0.2}_{-0.2}$ & 0.5$^{+0.2}_{-0.2}$ & 0.5$^{+0.2}_{-0.2}$ & 0.3$^{+0.2}_{-0.1}$ & 0.3$^{+0.2}_{-0.1}$ & 0.3$^{+0.2}_{-0.1}$ & see text\\
Quadratic LD coefficient & u$_2$ &  & -0.1$^{+0.1}_{-0.2}$ & -0.1$^{+0.1}_{-0.2}$ & -0.1$^{+0.1}_{-0.2}$ & 0.2$^{+0.2}_{-0.1}$ & 0.2$^{+0.2}_{-0.2}$ & 0.2$^{+0.2}_{-0.1}$ & see text\\
\hline
\hline
Derived parameters &  &  &  & & & & & \\
\hline
Planetary radius & R$_{\rm p}$ & R$_{\rm Jup}$ & 0.74$^{+0.05}_{-0.05}$ & 0.73$^{+0.05}_{-0.05}$ & 0.73$^{+0.05}_{-0.05}$ & 0.98$^{+0.06}_{-0.06}$ & 0.97$^{+0.06}_{-0.06}$ & 0.98$^{+0.06}_{-0.06}$\\
Scaled semi-major axis & $a/R_\star$ &  & 45$^{+7}_{-5}$ & 44$^{+2}_{-2}$ & 44$^{+2}_{-2}$ & 52$^{+9}_{-7}$ & 42$^{+2}_{-2}$ & 44$^{+2}_{-2}$\\
Orbital inclination & i & deg & 89.34$^{+0.08}_{-0.09}$ & 89.33$^{+0.10}_{-0.10}$ & 89.4$^{+0.3}_{-0.1}$ & 89.34$^{+0.08}_{-0.09}$ & 89.3$^{+0.1}_{-0.1}$ & 89.25$^{+0.09}_{-0.10}$\\
Transit duration & $T_{14}$ & hr & 7.21$^{+0.09}_{-0.08}$ & 7.20$^{+0.07}_{-0.07}$ & 7.17$^{+0.10}_{-0.09}$ & 7.6$^{+0.1}_{-0.1}$ & 7.53$^{+0.10}_{-0.10}$ & 7.6$^{+0.1}_{-0.1}$\\
\hline
Bayesian Information Criterion & BIC &  & -43314$^{}_{}$ & -43322$^{}_{}$ & -43306$^{}_{}$ & -69515$^{}_{}$ & -69523$^{}_{}$ & -69506$^{}_{}$\\
\hline
\end{tabular}
\tablefoot{
        \tablefoottext{a}{The best-fit values are given as the median of the posterior distributions, and the uncertainties are given as the $16^{\rm th}$ and $84^{\rm th}$ percentiles.}
        \tablefoottext{b}{We adopted the stellar mass and radius derived by \cite{Suarez_2022NatAs...6..232S}: $M=1.170\pm 0.060$ M$_{\odot}$, $R=1.278\pm0.070$ R$_{\odot}$}
        \tablefoottext{c}{Referred to the stellar descending point.}

}
\end{sidewaystable*}

\begin{sidewaystable*}
\small
\caption{Model parameters for the fit of the CHEOPS data.\tablefootmark{a}}\label{tab:cheopsFit}
\renewcommand{\arraystretch}{1.5} 
\begin{tabular}{lll|cccc|l}
\hline\hline
Jump parameters & Symbol & Units & no planet & free $P_{\rm orb}$ & $P_{\rm orb}=45.0033$ d & $P_{\rm orb}=45.0033$ d & Prior\\
                &        &       &    &  circular orbit & circular orbit & eccentric orbit\\
\hline
\textit{SHO kernel} \\
Amplitude scale & $\log\sigma_{\rm SHO}$ &  & -2.9$^{+0.2}_{-0.2}$ & -2.9$^{+0.2}_{-0.2}$ & -2.9$^{+0.2}_{-0.2}$ & -2.9$^{+0.2}_{-0.2}$ & $\mathcal{U}$(-6,-2)\\
Primary frequency$\equiv$Rotational frequency  & $\nu_{\rm 0}\equiv \nu_{\rm rot}$ & day$^{-1}$ & 0.345$^{+0.002}_{-0.002}$ & 0.345$^{+0.002}_{-0.002}$ & 0.345$^{+0.002}_{-0.002}$ & 0.345$^{+0.002}_{-0.002}$ & $\mathcal{N}$(0.346,0.002)\\
Timescale & $\log \frac{l_{\rm SHO}}{1~day}$ &  & 0.8$^{+0.5}_{-0.4}$ & 1.2$^{+0.5}_{-0.4}$ & 1.2$^{+0.5}_{-0.4}$ & 1.1$^{+0.5}_{-0.4}$ & $\mathcal{U}$(0,5)\\
\textit{Instrument-related parameters} \\
Decorrelation against $x_{\rm PSF}$ & $c_{\rm x}$ & pix$^{-1}$ & 0.00006$^{+0.00003}_{-0.00003}$ & 0.00006$^{+0.00003}_{-0.00003}$ & 0.00006$^{+0.00003}_{-0.00003}$ & 0.00006$^{+0.00003}_{-0.00003}$ & $\mathcal{U}$(-0.0005,0.0005)\\
Decorrelation against $y_{\rm PSF}$ & $c_{\rm y}$ & pix$^{-1}$ & -0.00040$^{+0.00003}_{-0.00003}$ & -0.00040$^{+0.00003}_{-0.00003}$ & -0.00040$^{+0.00003}_{-0.00003}$ & -0.00040$^{+0.00003}_{-0.00003}$ & $\mathcal{U}$(-0.0005,0.0005)\\
\textit{Stellar and planetary parameters} \\
Stellar density & $\rho_\star$ & $\rho_\sun$  & -  & 0.55$^{+0.07}_{-0.06}$ & 0.43$^{+0.01}_{-0.01}$ & 0.56$^{+0.09}_{-0.08}$ & $\mathcal{N}$(0.56,0.10)\\
Time of inferior conjunction - 2\,450\,000 & $T_0$ & BJD$_{\rm TDB}$ & -  & 9931.582$^{+0.001}_{-0.001}$ & 9931.583$^{+0.001}_{-0.001}$ & 9931.583$^{+0.001}_{-0.001}$ & $\mathcal{U}$(9901.8,9932.5)\\
Orbital period & P$_{\rm orb}$ & day & -  & 58$^{+7}_{-7}$ & -  & - & $\mathcal{U}$(35,100) \\
Planet-to-star radius ratio & $R_p/R_\star$ &  & -  & 0.060$^{+0.005}_{-0.005}$ & 0.059$^{+0.005}_{-0.005}$ & 0.059$^{+0.005}_{-0.006}$ & $\mathcal{U}$(0,0.2)\\
Impact parameter & $b$ &  & -  & 0.11$^{+0.09}_{-0.07}$ & 0.09$^{+0.07}_{-0.06}$ & 0.12$^{+0.11}_{-0.08}$ & $\mathcal{U}$(9901.8,9932.5)\\
Orbital eccentricity & $e$ &  & -  & -$^{}_{}$ & -$^{}_{}$ & 0.12$^{+0.09}_{-0.06}$ & $\mathcal{U}$(0,1)\\
Argument of periastron\tablefootmark{c} & $\omega_\star$ & rad & -  & -$^{}_{}$ & -$^{}_{}$ & -1$^{+2}_{-1}$ & $\mathcal{U}$(0,2$\pi$)\\
Linear LD coefficient & u$_1$ &  & -  & 0.3$^{+0.2}_{-0.2}$ & 0.3$^{+0.2}_{-0.2}$ & 0.3$^{+0.2}_{-0.2}$ & see text\\
Quadratic LD coefficient & u$_2$ &  & -  & 0.1$^{+0.2}_{-0.2}$ & 0.1$^{+0.2}_{-0.2}$ & 0.1$^{+0.2}_{-0.1}$ & see text\\
\hline
\hline
Derived parameters &  &  &  & & & \\
\hline
Planetary radius & R$_{\rm p}$ & R$_{\rm Jup}$ & -  & 0.74$^{+0.07}_{-0.07}$ & 0.74$^{+0.07}_{-0.07}$ & 0.72$^{+0.08}_{-0.09}$\\
Scaled semi-major axis & $a/R_\star$ &  & -  & 52$^{+6}_{-6}$ & 40.3$^{+0.3}_{-0.4}$ & 44$^{+2}_{-2}$\\
Orbital inclination & i & deg & -  & 89.88$^{+0.08}_{-0.10}$ & 89.88$^{+0.08}_{-0.11}$ & 89.8$^{+0.1}_{-0.2}$\\
Transit duration & $T_{14}$ & hr & -  & 9.02$^{+0.09}_{-0.09}$ & 9.01$^{+0.08}_{-0.09}$ & 9.00$^{+0.10}_{-1.90}$\\
\hline
Bayesian Information Criterion & BIC &  & -34048$^{}_{}$ & -34064$^{}_{}$ & -34072$^{}_{}$ & -34054$^{}_{}$\\
\hline
\end{tabular}
\tablefoot{
        \tablefoottext{a}{The best-fit values are given as the median of the posterior distributions, and the uncertainties are given as the $16^{\rm th}$ and $84^{\rm th}$ percentiles.}
        \tablefoottext{b}{We adopted the stellar mass and radius derived by \cite{Suarez_2022NatAs...6..232S}: $M=1.170\pm 0.060$ M$_{\odot}$, $R=1.278\pm0.070$ R$_{\odot}$}
        \tablefoottext{c}{Referred to the stellar descending point.}
}
\end{sidewaystable*}

\subsection{Ground-based follow-up} \label{sec:groundbasedfollowup}

\begin{figure}
    \centering
    \includegraphics[width=0.5\textwidth]{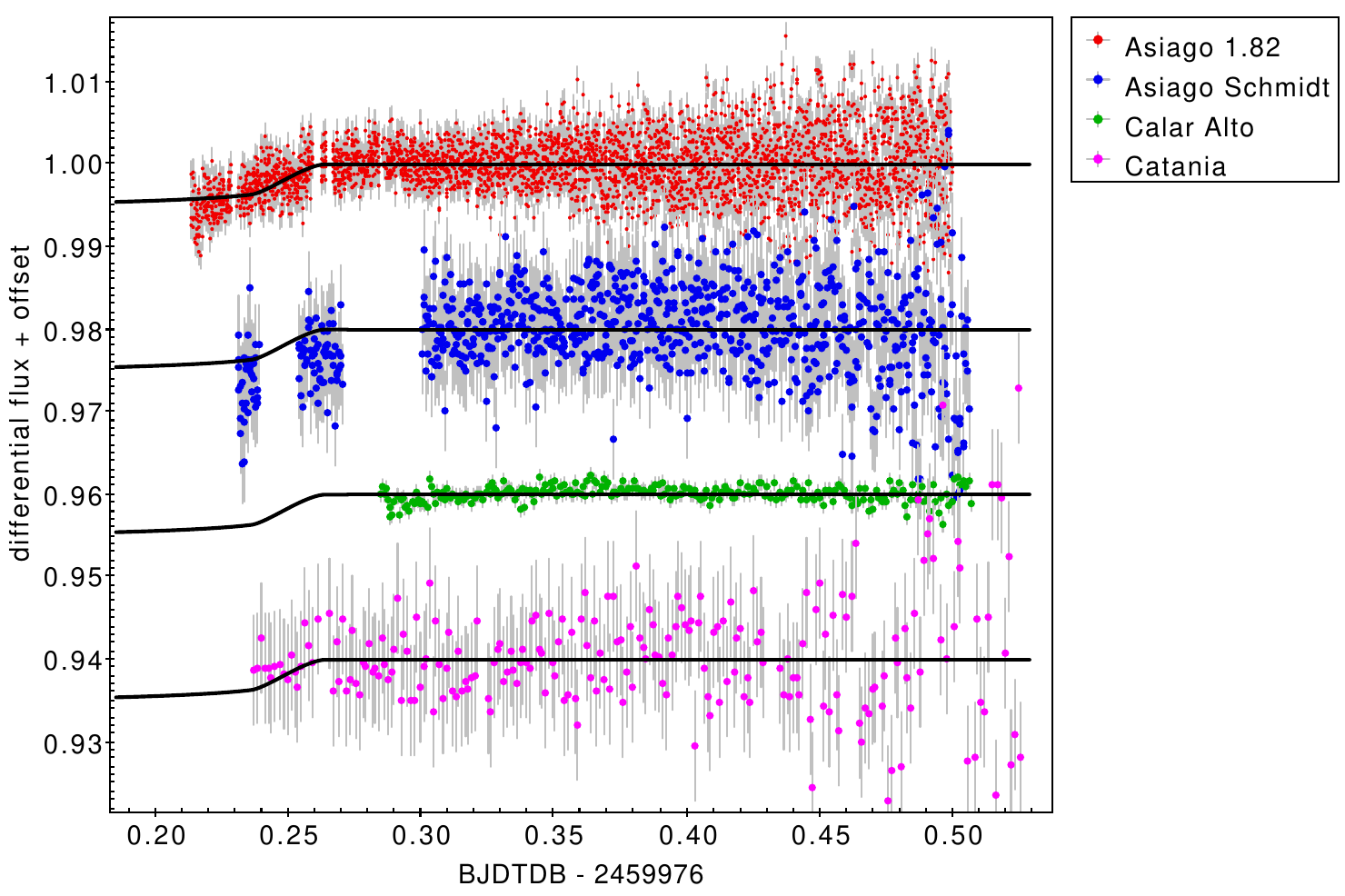}
    \caption{Ground-based photometry collected on the night 31 January - 01 February 2023 from different European Observatories.}
    \label{fig:groundphotometry}
\end{figure}

All the ground-based photometric data we gathered during the night of 31 January 2023 (Sect.~\ref{sec:ground}) are plotted in Fig.~\ref{fig:groundphotometry}. We remark that all the instruments that we used could in principle significantly detect a transit with a depth of $\sim 0.5\%$. We globally fitted all of them with the \texttt{PyORBIT} code \citep{Malavolta2016,Malavolta2018} (using the \texttt{emcee} Markov Chain Monte Carlo sampler) to search for a transit of planet $e$ that was predicted to occur by assuming the orbital period $P_e=45.0003$~d. Except for the shallow transit of V1298\,Tau\,c (ingress: 2459976.28146; centre: 2\,459\,976.37856; egress: 2\,459\,976.47566 BJD$_{\rm TDB}$, based on the ephemeris of \citealt{Feinstein_2022ApJ...925L...2F}; depth $\sim 0.1\%$) no other transit event was predicted in that observing window. To model the observations we adopted a single-planet transit model where all the stellar and planetary parameters were fixed to the values published by \citet{Feinstein_2022ApJ...925L...2F}, and only the central transit time $T_0$ and a linear baseline $f(t) = a_0+a_1\cdot t$ were left free to vary using a uninformative prior. After $100\,000$ steps with a thinning factor of 100 and a burn-in phase of $20\,000$ steps, the posterior distribution of $T_0$ regularly converged to $T_{0} = 2\,459\,976.0918 \pm 0.0016 $~BJD$_{\rm TDB}$ meaning that just the egress of planet $e$ was captured by the Asiago Copernico and Schmidt \acp{LC} and missed by the other two observing sites, as can be seen by the maximum a-posteriori probability (MAP) model plotted as a black line in Fig.~\ref{fig:groundphotometry}. This unexpected finding is in striking contrast with our initial prediction $T_{0} = 2\,459\,976.8309$ BJD$_{\rm TDB}$ and would imply an $O-C$ of about $-12$~hours.

We note that the egress-like feature falls at the very beginning of our coverage, during nautical twilight at Asiago, and this feature could be influenced by the sky conditions. On the other hand, all the instrumental diagnostics look perfectly nominal for both \acp{LC}, which agree equally well with the best-fit transit model, and in particular with the transit depth and the ingress/egress duration ($T_{14}$) of planet V1298\,Tau\,$e$. 

\section{Discussion and Conclusions} \label{sec:conclusions}

The main result presented in this work is the detection of a transit in the LC of the young star V1298\,Tau collected by CHEOPS between 17 and 18 December 2022. The signal falls within a time interval when a transit of the planet V1298\,Tau\,$e$ was predicted to occur, if its orbital period were close to $P_e=45.0033$ days. This result looks not compatible with an orbital period close to $P_e=46.7681$ days, which is the most probable value proposed by \cite{sikora2023}, because no transit would then be expected to fall within the third CHEOPS observing slot, even though this conclusion is necessarily based on a very limited amount of information that we have so far about the orbital dynamics of V1298\,Tau\,$e$. 

We reanalysed the individual transits observed by $Kepler/K2$ and TESS that have been ascribed to V1298\,Tau\,$e$, and compared them with the signal that we detected in the CHEOPS LC, finding some differences which make a coherent description of this puzzling system challenging. 
The main observational results and findings from the comparative analysis of the three $Kepler/K2$, TESS, and CHEOPS transits can be summarised as follows (see Section \ref{sec:cheopslcanalysis} for details):
\begin{itemize}
    \item the transit duration as seen by CHEOPS is significantly longer than that derived from $Kepler/K2$ and TESS observations. We note that our derived transit duration from TESS data is longer than that observed by $Kepler/K2$, but compatible within $\sim3\sigma$;
    \item without constraining the orbital period in the MCMC analysis, we get posteriors for $P_{\rm orb}$ that are in agreement within $1\sigma$ for the three telescopes, but the impact parameter derived from CHEOPS data is significantly lower ($b=0.11^{+0.09}_{-0.07}$), as it is expected for a transit with a longer duration;
    \item the transit depth as seen by CHEOPS is consistent with that of the $Kepler/K2$ transit, while the transit depth measured from the TESS LC is larger. This implies a radius for the transiting companion of $\sim0.74$ $R_{\rm Jup}$ ($Kepler/K2$ and CHEOPS), or $\sim0.98$ $R_{\rm Jup}$ (TESS). We remark that \cite{Feinstein_2022ApJ...925L...2F} found a radius $\sim3\sigma$ larger in the TESS data ($R_p=0.89\pm0.04$ $R_{\rm Jup}$) than that found in the $Kepler/K2$ data by \cite{David_2019ApJ...885L..12D}, while in our case this discrepancy increases to $\sim4\sigma$ (i.e. the ratio $R_p/R_\star$ measured by TESS is $\sim34\%$ higher than for $Kepler/K2$). That is because our derived TESS transit is deeper than that of \cite{Feinstein_2022ApJ...925L...2F}, as shown in Fig. \ref{fig:transits}, implying a larger planet radius ($0.98\pm0.06$ vs. $0.89\pm0.04$ $R_{\rm Jup}$). 
\end{itemize}

Regarding the comparison between our derived TESS LC and that of \cite{Feinstein_2022ApJ...925L...2F}, we investigated whether a similar difference in the transit depths is also observed for planet $b$, which has a transit depth comparable to that of planet $e$ (Appendix \ref{sec:comparisonplanetb}). In this case, we recover a transit depth slightly shallower, but well in agreement within the error bars (Fig. \ref{fig:planetb}), and conclude that apparently a significant difference with \cite{Feinstein_2022ApJ...925L...2F} is only observed for planet $e$. 

These results do not have a straightforward interpretation, and bring into question even the identity of the transiting companion(s). Nonetheless, it must be emphasised that any reasonable explanation is naturally constrained by the limited number of the observed transits. We propose two main interpretative frameworks based on the available data. 

\begin{figure}
    \centering
    \includegraphics[width=\linewidth]{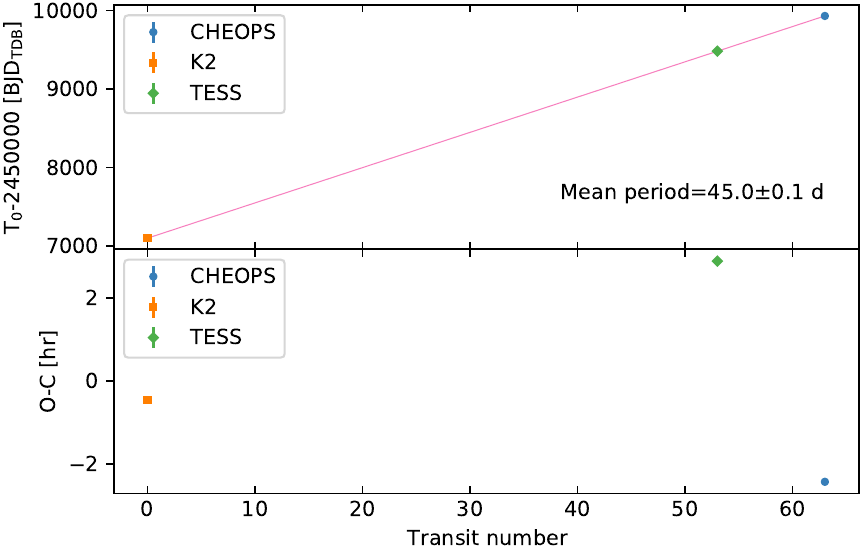}
    \caption{Transit timing variations for the three transits observed by $Kepler/K2$, TESS, and CHEOPS assuming that they are all ascribable to V1298\,Tau\,$e$. The observed epochs of central transit have been fitted with a linear function, resulting in a mean orbital period of $45.0\pm0.1$ days.  }
    \label{fig:TTV}
\end{figure}

First, let us make the hypothesis that $Kepler/K2$, TESS, and CHEOPS indeed observed the same planet V1298\,Tau\,$e$, as previously identified by \cite{David_2019ApJ...885L..12D} and \cite{Feinstein_2022ApJ...925L...2F}. Thanks to CHEOPS, this assumption would solve the uncertainty on the orbital period, implying that $P_e\sim45$ days, and also implying that the outermost planet in the system experiences changes to the orbital parameters on a timescale of a few years. Indeed, a change in the orbital elements occurring over short timescales would explain the presence of large TTVs (Fig. \ref{fig:TTV}), and transit duration variations (TDV). As an illustrative case, we performed a few N-body numerical integration of the V1298\,Tau system for different random choices of the initial orbital angles of planet $b$ and $e$ (mean anomaly and nodal longitude). For this exercise, we adopted the masses of V1298\,Tau\,$b$ and $e$ derived by \cite{Suarez_2022NatAs...6..232S}, and mild eccentricities (0.01--0.05). We see from Fig. \ref{fig:dynam_integr} that the orbital period could change significantly over a very short timescale, therefore the orbit of V1298\,Tau\,$e$ can be likely chaotic. Changes in the orbital parameters can determine a corresponding change in the transit parameters over a few years, which is an interval similar to the time span covered by the photometric observations. We emphasise that the result of Fig. \ref{fig:dynam_integr} is only suggestive because of our limited prior knowledge about the system's architecture and fundamental properties. As an alternative and illustrative case, we ran another loosely constrained set of dynamical integration over a time interval of 10 years by imposing circular orbits, and fixing the longitude of the ascending nodes to 180 deg. In this case, planet masses have been drawn from the posteriors derived by \cite{sikora2023}. The results show that there could be a significant probability that we may have missed transits in the first two CHEOPS observing slots, if the orbital period $P_e$ is close to 46.76 and 44.17 days, respectively, and TTVs are $\lesssim$1 hour. If that is the case, this would imply that the observed transit in the third CHEOPS slot belongs to a fifth companion. Any in-depth analysis of the system’s dynamics can be performed only after additional transits of planet $e$ will be observed, and must be left to future studies.
We fitted the three transits by fixing $P_e$ to 45.0033 days, and tested both circular and eccentric orbits (Tab. \ref{tab:k2TessFit} and \ref{tab:cheopsFit}). In all the cases, the BIC statistics does not strongly favour these models over those with $P_e$ treated as a free parameter. CHEOPS data are better fitted by a circular model ($\Delta BIC=-18$), but the posterior of the stellar density significantly differs from the prior. That is not surprising, because a larger stellar radius (lower stellar density) has to be assumed to explain a longer transit duration when fixing $P_e$ to 45 days. We recover a correct stellar density when we include the eccentricity as a free parameter, and we get $e_e=0.12^{+0.09}_{-0.06}$. In the cases of $Kepler/K2$ and TESS, a circular model does not result in a similar issue. This first scenario is tentatively supported by ground-based photometry (Sect. \ref{sec:groundbasedfollowup}) that we collected 45 days after the transit detected by CHEOPS. The results from this follow-up do not exclude the possibility that we detected the egress of V1298\,Tau\,$e$, with the transit occurring $\sim-12$ hr earlier than predicted assuming the reference period from \cite{Feinstein_2022ApJ...925L...2F}. We note that a study of the V1298\,Tau system focused on the determination of TTVs is not yet available in the literature. Obviously, to confirm this scenario additional and intensive photometric follow-up is necessary. Unfortunately, at the time of writing, we know that TESS is not going to observe V1298\,Tau during Cycle 6 starting from Sept. 2023\footnote{\url{https://heasarc.gsfc.nasa.gov/wsgi-scripts/TESS/TESS-point_Web_Tool/TESS-point_Web_Tool/wtv_v2.0.py/}}. New observations with CHEOPS are indeed required, and they must be scheduled allowing for a sufficiently large observational window to take into account several hours of TTVs which are presently not predictable.  

In a second framework, we interpret the transit detected by CHEOPS as due to a new planet-sized companion in the system, different from that observed by $Kepler/K2$ and TESS, which here we assume to be produced by the same planet. Assuming circular orbits, this would reconcile the different observed transit duration, and it would exclude $P_e=45.0033$ days from the list of possible orbital periods of the $Kepler/K2$ and TESS companion. Future photometric follow-up is indeed necessary also to confirm this scenario, in case no additional transits corresponding to $P\sim 45$ days will be detected. However, taking into account the 71-day time span of almost uninterrupted observations, the chances that $Kepler/K2$ would have detected one transit of a planet with $P_{\rm orb}=58\pm7$ days are not negligible, in that it falls nearly 2$\sigma$ within the duration of the $Kepler/K2$ observing window.  

Whatever the correct scenario is, we need to find an explanation also for the different transit depth observed by $Kepler/K2$, TESS, and CHEOPS. For a preliminary assessment, we neglect the differences in the $Kepler/K2$ (CHEOPS) and TESS passbands (Fig. \ref{fig:passbands}). In active stars where a significant fraction of the disk is covered by spots, as it is expected in the case of V1298\,Tau, unocculted starspots during a transit would produce an apparent increase in the planet's radius up to $10\%$, when the transits occur at very different locations in a starspot modulated LC (e.g. see Eq. 17 in \citealt{Morris2020ApJ...893...67M}): assuming that the planet does not cover any of the spots during its transit, when the unocculted spots occupy a smaller area of the stellar disc (i.e. at a maximum of the LC) the out-of-transit stellar brightness baseline will be higher, and the transit depth will be shallower, and the other way round. Here, the transits observed by $Kepler/K2$ and CHEOPS occur close to a minimum in their LCs, while the TESS transit is located close to a maximum of the stellar brightness (Fig. \ref{fig:LCs}). Thus, assuming that the results of a GP model do not depend on where the transit is located on the LC, we would expect the TESS transit to be shallower than the other two, while we observe the opposite. We could reconcile the observations of the three individual transits, at least qualitatively, by assuming that the atmosphere and surface magnetic activity of V1298\,Tau is faculae-dominated instead of being spot-dominated. For pre-main sequence stars, this is a plausible scenario, especially for fast rotators for which an anti-correlation between X-ray and simultaneous optical variability is observed (e.g. \citealt{guarcello2019A&A...628A..74G}). Nonetheless, that possibility has not been yet investigated for V1298\,Tau. We also investigated a possible dependence of the transit depths of planet $b$ from the specific location of the transit signals on the TESS LC. The two transits occurred at different phases of the LC, and they perfectly match (Fig. \ref{fig:transits_b_tess}).

Considering the effects of starspot occultation by a transiting planet, this might in principle explain the observed differences if we assume an uninterrupted band of dark spots extended in latitude running through all the stellar disk. When the transit chord runs within that band, depending on the fraction of the dark strip covered by the planet disk a transit depth would be more or less affected, and significantly change. This is indeed a \textit{ad hoc} scenario that should be tested with additional transit observations and modelling, but it would explain the observations if we assume that the planet was covering a larger part of the dark band when observed $K2/Kepler$ and CHEOPS.
Starspots on the stellar disc may produce small variations in the apparent duration of a transit (longer or shorter), of the order of 4\% as estimated by \cite{Oshagh2013A&A...556A..19O}. Therefore, it looks hard to invoke starspots to explain the 1.5-hour longer duration observed by CHEOPS, which is nearly 20\% longer than the duration of the $Kepler/K2$ and TESS transits.

The detection of a new transit-like signal with interesting properties is indeed an important outcome of this study, nonetheless our result does not allow us to draw unambiguous conclusions on the identity of the transiting companion(s) detected by three different telescopes. That is not surprising in that V1298\,Tau is a very challenging system to be characterised, both in terms of planet properties and architecture, and stellar activity. The more plausible interpretation may be that we observed the same transiting planet detected by $Kepler/K2$ and TESS with an orbital period of $\sim 45$ days (from a linear ephemeris), which shows TTVs and TDVs due to dynamical instability of its orbit. After all, we detected a transit within the time window where it was expected to occur assuming one of the orbital periods in the grid of \cite{Feinstein_2022ApJ...925L...2F} as a guidance. The hypothesis of a fifth planet in the system appears less probable with the data currently available. We plan to observe V1298 Tau again with the CHEOPS telescope in the near future, with the primary aim of detecting more transit signals of planet $e$, and revising the results of this study. The priority will be looking for transits to confirm an orbital period of $\sim$45 days. At the same time, we will analyse a time series of hundreds of RVs of V1298~Tau, most of them collected with the HARPS-N spectrograph as a continuation of the work by \cite{Suarez_2022NatAs...6..232S}. As also shown by the studies of \cite{sikora2023} and \cite{blunt2023AJ....166...62B}, the analysis of the RVs for such an active star is notoriously very challenging, especially if the transit ephemeris are not well constrained for all the planets in the system. Our goal will be to exploiting the synergy between CHEOPS and RV follow-up in order to improve the characterisation of the orbital and fundamental physical parameters of the planets in the V1298~Tau system. This, in turn, will allow for a well-informed analysis of the planet formation history, and of the current architecture and dynamical state of the system, improving the results of a preliminary investigation carried out by \cite{turrini2023arXiv230708653T}.

\begin{acknowledgements}
This work has been supported by the PRIN-INAF 2019 "Planetary systems at young ages (PLATEA)" and ASI-INAF agreement n.2018-16-HH.0. CHEOPS is an ESA mission in partnership with Switzerland with important contributions to the payload and the ground segment from Austria, Belgium, france, Germany, Hungary, Italy, Portugal, Spain, Sweden and the United Kingdom. GSca, VNa, GBr, and LBo acknowledge support from CHEOPS ASI-INAF agreement n. 2019-29-HH.0. L.\,M. acknowledges support from the ``Fondi di Ricerca Scientifica d'Ateneo 2021'' of the University of Rome ``Tor Vergata''. A.S.M. acknowledges financial support from the Spanish MICINN project PID2020-117493GB-I00 and the Government of the Canary Islands project ProID2020010129. VBe, MMa, NLo, and MZa acknowledge support from the Agencia Estatal de Investigaci\'on del Ministerio de Ciencia e Innovaci\'on under grants PID2019-109522GBC51(53). This work includes data collected with the TESS mission, obtained from
the MAST data archive at the Space Telescope Science Institute (STScI). Funding for the TESS mission is provided by the NASA Explorer Program. STScI is operated by the Association of Universities for Research in Astronomy, Inc., under NASA contract NAS 5–26555. This paper also includes data collected by the Kepler mission and obtained from the MAST data archive at the Space Telescope Science Institute (STScI). Funding for the Kepler mission is provided by the NASA Science Mission Directorate. STScI is operated by the Association of Universities for Research in Astronomy, Inc., under NASA contract NAS 5–26555.
The results presented in this work are also based on observations collected at the Centro Astron\'{o}mico Hispano en Andalucía (CAHA) at Calar Alto, operated jointly by Junta de
Andalucía and Consejo Superior de Investigaciones Cient\'{i}ficas (IAA-CSIC), and at Copernico 1.82m and Schmidt 67/92 telescopes (Asiago Mount Ekar, Italy), INAF - Osservatorio Astronomico di Padova. We kindly thank S.~Ciroi for having shared his observing time at the Copernico 1.82m telescope. This work is dedicated to the memory of A.C.P., a young star returned too soon among the stars.
\end{acknowledgements}

%
%

\bibliographystyle{aa}
\bibliography{references}

\begin{appendix}

\section{GP modelling of the light curves}\label{sec:activity}

\begin{figure*}
    \centering
    \includegraphics[width=.49\linewidth]{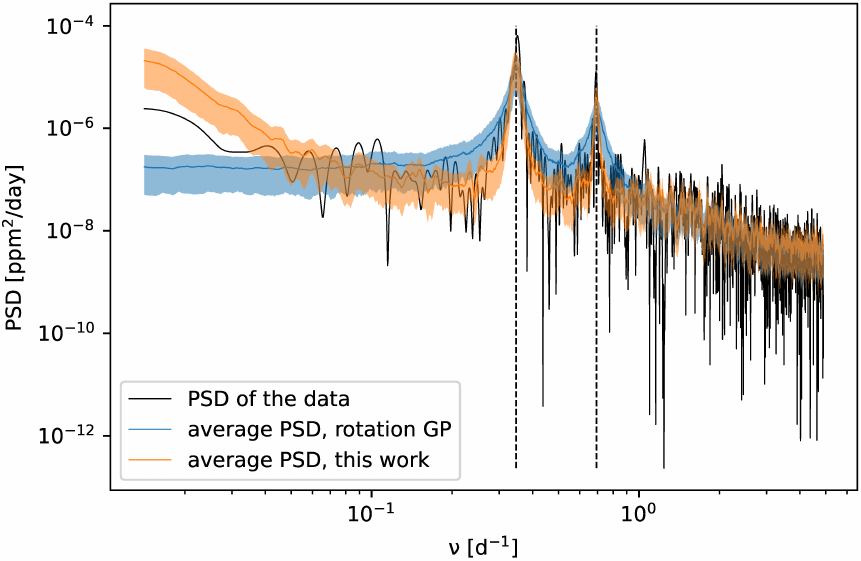}
    \includegraphics[width=.49\linewidth]{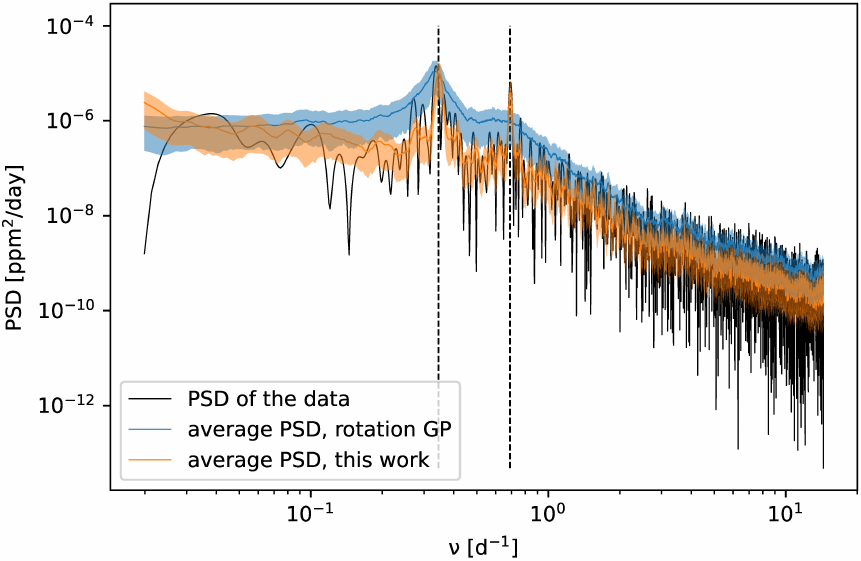}
    \caption{Analysis of the PSD of the $Kepler/K2$ (left) and TESS (right) LCs. In each panel, the black solid line shows the PSD of the LC after masking the planetary transits. The two vertical dashed lines mark the frequencies $0.35~d^{-1}$ and $0.70~d^{-1}$. The confidence band in orange shows the average periodogram obtained using the rotation GP, as discussed in the text. Similarly, the blue confidence band corresponds to the GP model used in this work.}\label{fig:periodograms}
\end{figure*}

The $Kepler/K2$ and TESS \acp{LC} of V1298~Tau clearly show a periodic signal that gradually changes with time. This kind of correlated noise is typically generated by stellar active regions co-rotating with the star. To extract the characteristic frequencies of the stellar signal, we first masked the planetary transits in the $Kepler/K2$ and TESS \acp{LC} by using the ephemeris obtained by \citet{David_2019ApJ...885L..12D} and \citet{Feinstein_2022ApJ...925L...2F}. Then we computed the \ac{PSD} of the remaining \acp{LC} using \ac{LS} periodogram \citep{LS1982}. Both periodograms show two peaks at frequencies $\nu\sim0.35~d^{-1}$ and $\nu\sim0.70~d^{-1}$ respectively (Fig.\ref{fig:periodograms}). Since the latter is twice the former, we speculate that the peak at lower frequency corresponds to the stellar rotation ($P_{\rm rot}\sim$2.9 d) while the second peak is its first harmonic.

In their analysis of the planetary transits, \citet{David_2019ApJ...885L..12D} and \citet{Feinstein_2022ApJ...925L...2F} modelled the stellar variability as a \lq\lq rotation\rq\rq\ \ac{GP} defined as the sum of two \ac{SHO} kernels. The simple \ac{SHO} kernel has a \ac{PSD} of the form:
\begin{equation}
    S(\nu)=\sqrt{\frac{2}{\pi}}\frac{S_0\nu^4_0}{\left(\nu^2-\nu^2_0\right)^2+\nu^2\nu^2_0/Q^2},
\end{equation}
where $\nu_0$ is the undamped frequency of the oscillator (not to confuse with the corresponding angular frequency $\omega_0$), $S_0$ fixes the scale of the \ac{PSD} and $Q$ is the quality factor of the oscillations. The simple \ac{SHO} kernel can model only one peak in the \ac{PSD} of the data. In the case of pluri-modal \acp{PSD}, a \ac{SHO} for each mode in the \ac{PSD} is needed. This explains the approach of \citet{David_2019ApJ...885L..12D} and \citet{Feinstein_2022ApJ...925L...2F}.

The exact parametrization of the mixture of \ac{SHO} kernels is discussed in \citet{David_2019ApJ...885L..12D}. Since it is difficult to relate the mathematical representation of this \ac{GP} (scale $S$ and quality $Q$) to the physical properties of the \acp{LC} (amplitude of the correlated noise and its timescale), we converted the best-fit \ac{GP} parameters into physical via the equations\footnote{\url{https://celerite2.readthedocs.io/en/latest/api/python/#celerite2.terms.SHOTerm}}:
\begin{align}
    P_{\rm SHO}=&1/\nu_0,\label{eq:psho}\\
    \lambda_{\rm SHO}=&4\pi Q/\nu_0,\label{eq:lsho}\\
    \sigma_{\rm SHO}=&\sqrt{2\pi S_0\nu_0Q}.
\end{align}
For the case of the $Kepler/K2$ LC, \citet{David_2019ApJ...885L..12D} derived two \ac{SHO} \acp{GP} with periods of 2.87 d and 5.74 d, amplitudes of 12 mmag and 7 mmag and timescales of 14 d and 5 d. For the TESS \ac{LC}, \citet{Feinstein_2022ApJ...925L...2F} used the same rotation \ac{GP} model and they obtained periods of 2.97 d and 5.94 d and timescales of 6 d and 1 d. Unfortunately, they forgot to report the scales $S$ of the two \acp{GP}, which are needed to compute the corresponding noise amplitudes. In the following discussion, we thus assume the same amplitudes derived by \citet{David_2019ApJ...885L..12D}.

In both cases, the stellar rotation period is successfully recovered as the periodicity of the first \ac{SHO} component, the second component corresponding to the first harmonic. We remark though that the timescales of the rotation \ac{GP} are a few times the period of the oscillations (2.87 d and 2.97 d for $Kepler/K2$ and TESS respectively). This is critical in particular for the TESS analysis, where the first \ac{SHO} component has a timescale of roughly 2 oscillation periods, while the second component has a timescale that comparable with or shorter than the stellar rotation period. From a physical point of view, when the timescale of the oscillation is close or less than the period of the oscillations, then the periodicity of the signal is suppressed by the fast exponential decay due to the short timescale. As a result, such systems do show a correlated signal but without the characteristic repeating pattern similar to undamped oscillators, clearly visible in the $Kepler/K2$ and TESS \acp{LC} (Fig.~\ref{fig:LCs}).

From a more quantitative perspective, we compared the \ac{PSD} of the data with the typical \ac{PSD} corresponding to the \acp{GP} derived by \citet{David_2019ApJ...885L..12D} and \citet{Feinstein_2022ApJ...925L...2F}. For each \ac{GP} we generated 1000 \acp{LC} using the same timestamps of the corresponding \ac{LC} (either $Kepler/K2$ or TESS) to take into account any aliasing effect due to the time sampling. Finally, for each set of 1000 \acp{LS}, we computed the average \ac{LS} periodogram and its 1-$\sigma$ uncertainty. The results are shown in Fig.~\ref{fig:periodograms}. The most striking feature of these average periodograms is that they recover the two peaks in the periodogram of the data, but the peaks are broader than the peaks in the periodogram of the data. This means that the quality factor $Q$ of the oscillations in the rotation \acp{GP} underestimate the true values or, in physical units, the timescales of the oscillations are shorter than in the data. This is most critical for the TESS \ac{LC}, where the $Q$ of the second \ac{SHO} \ac{GP} is so low (and the corresponding timescale is so short) that the corresponding peak at $\nu\sim0.70~d^{-1}$ is completely flattened out.

A closer look at the periodograms shows a hint of a decreasing slope at low frequencies ($\nu<0.3~d^{-1}$). This may indicate the presence of an additional correlated noise (other than the stellar rotation) with a \ac{PSD} that decreases with $\nu$. The presence of a slope at low frequencies is a challenge for the \ac{SHO} \ac{GP} and also for the rotation \ac{GP}, as by construction their corresponding \ac{PSD} is flat for frequencies lower than the peak frequencies.

We ran a few simulation tests by generating \acp{LC} using a rotation \ac{GP} combined with an overdamped \ac{SHO}, whose \ac{PSD} monotonically decreases with $\nu$ \citep{Foreman2017}. In order to save computing time, we fitted the \acp{LC} by means of least-square regression. We found that the presence of a neglected \ac{GP} biases the fit of the rotation \ac{GP} towards lower quality factors and shorter timescales. In the most critical cases, the fit of the rotation \ac{GP} also diverged towards unrealistic timescales and/or amplitudes of the correlated noise. This can explain why \citet{David_2019ApJ...885L..12D} and \citet{Feinstein_2022ApJ...925L...2F} use informative priors in their analysis. In the frequency domain, the fit of the simulated \acp{LC} leads to a rotation \ac{GP} whose corresponding \ac{LS} periodogram has broader peaks compared to the periodogram of the fitted \ac{LC}. In other words, the best-fit rotation \ac{GP} tries to include the spectral power due to the extra \ac{GP} by broadening the rotation peaks.

In the domain of Fourier analysis, the assumption of a \ac{PSD} that does not match the observations leads to a biased spectral decomposition of the signal. It is difficult to asses how this translates in the \ac{GP} analysis framework, and any further detailed analysis is not the scope of this work. Nonetheless, we tested a different \ac{GP} for two reasons. First of all, we aimed at a better representation of the correlated noise in the $Kepler/K2$ and TESS \acp{LC} in order to put our analysis on more solid ground. Secondly, we wanted to asses how the planetary parameters are affected by the choice of a different \ac{GP} model.

We modelled the correlated noise using the \texttt{S+LEAF} \texttt{python} library \citep[version 2.1.2][]{Delisle2020,Delisle2022}. This library implements the \ac{ESP} \ac{GP}, an approximation of the quasi-periodic kernel, with the generic element of the kernel matrix $k(\Delta t)$ described by the equation:
\begin{equation}
    k(\Delta t)=\sigma^2_{\rm ESP}\exp\left(-\frac{\Delta t^2}{2\lambda^2_{\rm ESP}}-\frac{\sin^2\left(\pi\nu_{\rm ESP}\Delta t\right)}{2\eta^2_{\rm ESP}}\right),
\end{equation}
where $\sigma_{\rm ESP}$ is the standard deviation of the \ac{GP}, $\lambda_{\rm ESP}$ is its timescale, $\nu_{\rm ESP}$ is the rotational frequency ($\equiv \nu_{\rm rot}$ in Table \ref{tab:k2TessFit}), and $\eta_{\rm ESP}$ is the scale of the oscillations, which sets how the spectral power is distributed among the harmonics of the periodic signal. In the current implementation of this kernel it is necessary to set the number of harmonics to include through the $nharm$ keyword: since the \ac{PSD} of the data shows two predominant peaks (the fundamental and first harmonics), we manually set the $nharm=2$.

We initially tried to use this \ac{GP} to fit by least-square regression the same stellar \acp{LC} discussed above, but we did not find any satisfactory results, mainly because the fit did not converge to a realistic set of \ac{GP} parameters. Once again, some tests showed that a low-frequency power excess might prevent the \ac{ESP} \ac{GP} alone to properly model the correlated noise. We thus added an extra \ac{SHO} \ac{GP} in order to take into account the slope at low frequencies discussed above, obtaining a more meaningful set of \ac{GP} parameters.

First of all, for the $Kepler/K2$ \ac{LC} the \ac{ESP} \ac{GP} converged to $\sigma_{\rm ESP}=18$ mmag, $P_{\rm ESP}=\nu^{-1}_{\rm ESP}$=2.88 d and $\lambda_{\rm ESP}$=17 d, while for the TESS \ac{LC} we obtained $\sigma_{\rm ESP}=11$ mmag, $P_{\rm ESP}$=2.90 d and $\lambda_{\rm ESP}$=30 d. Thus, we recovered similar periodicities, timescales and amplitudes for the two \acp{LC}, small differences being likely due to differences in the bandpasses and/or different active stellar latitudes at the epoch of the observations. Most importantly, we derived a \ac{GP} timescale approximately an order of magnitude longer than the rotation period.

The \ac{SHO} \ac{GP} converged to $\sigma_{\rm SHO}$=1.7 (3.3) ppm, $\P_{\rm SHO}$=1 (2) d and $\lambda_{\rm SHO}$=0.4 (0.2) d for the $Kepler/K2$ (TESS) \ac{LC}. By means of Eqs.~\ref{eq:psho}-\ref{eq:lsho}, these parameters translate into $Q$=1.2 and $Q$=0.3 for the $Kepler/K2$ and TESS \ac{LC} respectively. These two quality factors are close to the critical damping value of $Q=1/\sqrt2$, below which the \ac{PSD} of the \ac{SHO} \ac{GP} becomes monotonically decreasing with $\nu$. From a physical point of view, this corresponds to an overdamped \ac{GP} that can be representative of low-frequencies stellar processes. As an example, the overdamped \ac{SHO} \ac{GP} with fixed $Q=1/2$ is commonly used to model the surface granulation of convective stars \citep{Harvey1985, Kallinger2014}.

Similarly to what we have done above, we used our best-fit composite \ac{GP} to simulate 1000 \acp{LC}, using either the time sampling of $Kepler/K2$ or TESS. Then we computed the mean \ac{LS} periodogram, obtaining the \acp{PSD} shown in Fig.~\ref{fig:periodograms}. Our \ac{GP} model is able to better reproduce the two peaks in the \ac{PSD} of the data and also provides a better match at low frequencies. For all the reasons discussed so far in this Appendix, in our combined analysis of the stellar correlated signal and the planetary transits (Sect.~\ref{sec:k2Analysis} and \ref{sec:tessAnalysis}) we adopted the mixture of a \ac{SHO} and a \ac{ESP} \ac{GP} models. We also remark that we used uninformative priors on all the \ac{GP} parameters (see Table~\ref{tab:k2TessFit}) without any issue in the convergence of the \ac{MCMC} chains.

\section{Comparing our TESS transit of V1298\,Tau\,$b$ with that by \cite{Feinstein_2022ApJ...925L...2F}} \label{sec:comparisonplanetb}
\begin{figure*}
    \centering
    \includegraphics[width=.49\linewidth]{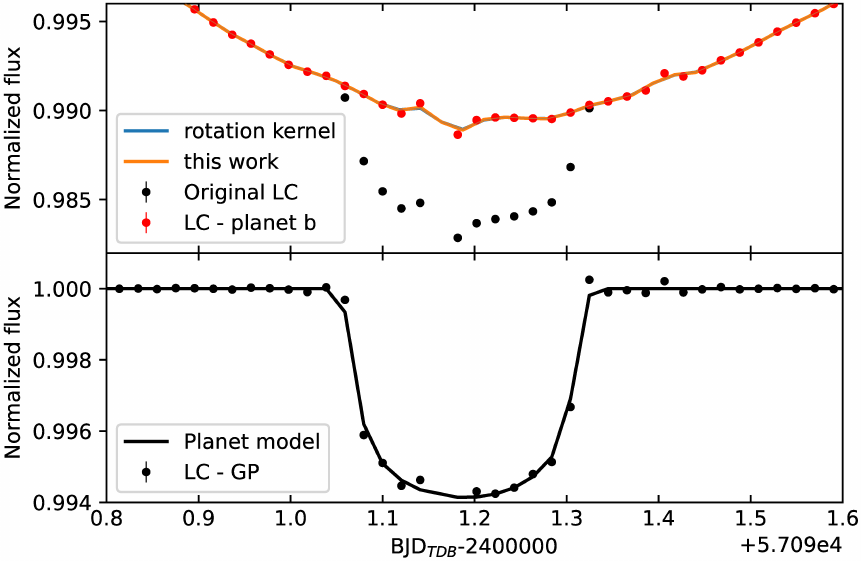}
    \includegraphics[width=.49\linewidth]{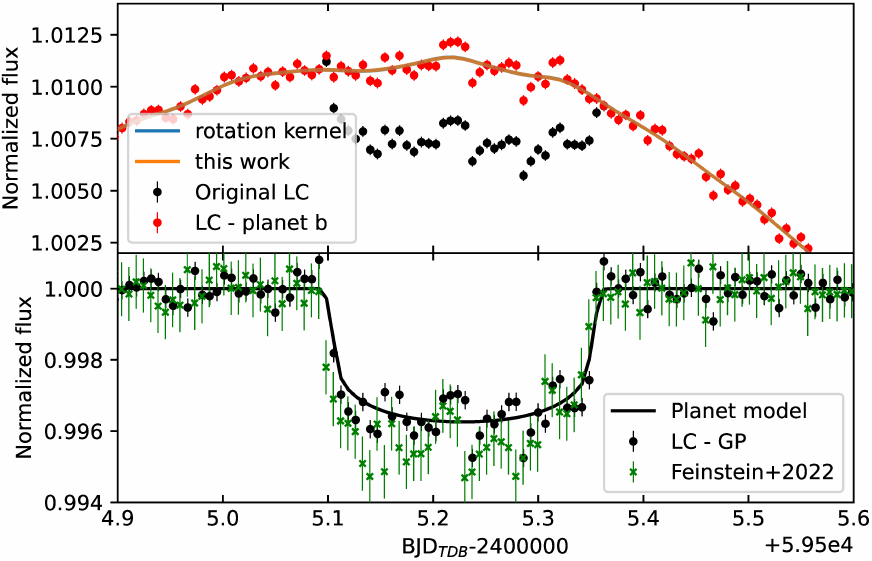}
    \caption{Portions of $Kepler/K2$ and TESS LCs showing a transit of planet V1298\,Tau\,$b$ (left and right panel respectively). The plots at the top show the undetrended LCs analysed in our work centred on the transit (black dots), and the data after removing the best-fit planetary transit (red dots) to show the GP signal during the transit timespan. The plots at the bottom show the transit signals after subtracting the best-fit \ac{GP} model. The corresponding detrended and flattened TESS light curve analysed by \cite{Feinstein_2022ApJ...925L...2F} is shown for comparison (green cross symbols). It has been obtained from the \texttt{jupyter} notebook made publicly available by \cite{Feinstein_2022ApJ...925L...2F} at https://github.com/afeinstein20/v1298tau$\_$tess/blob/main/notebooks/TESS$\_$V1298Tau.ipynb }\label{fig:planetb}
\end{figure*}

To assess whether and how our pipeline to extract the TESS LC (Sect. \ref{sec:datatess}) and \ac{GP} model affect the properties of the transits with respect to the analysis of \cite{Feinstein_2022ApJ...925L...2F}, we also considered transits of planet $b$. To simplify the analysis, we only focused on the transit around BJD 2\,457\,091.2 in the $Kepler/K2$ \ac{LC}, and the transit around BJD 2\,459\,505.2 in the TESS \ac{LC}. These transits do not overlap with transits of other planets, and are well sampled. The results of our \ac{LC} fitting are shown in Fig.~\ref{fig:planetb}, together with the best-fit rotational \ac{GP} signal derived by \citet{David_2019ApJ...885L..12D} and \citet{Feinstein_2022ApJ...925L...2F}. Our derived depth for the TESS transit is $R_p^2/R_\star^2$=0.0031$\pm$0.0003, which is shallower than the transit analysed by \cite{Feinstein_2022ApJ...925L...2F} ($R_p^2/R_\star^2$=0.0040$\pm$0.0002), with a discrepancy $\Delta(R_p^2/R_\star^2)/\sqrt{\Sigma(\sigma_{\rm R_p^2/R_\star^2}^2)}$=2.5. We confirm the difference in the transit depth as seen by the $Kepler/K2$ and TESS telescopes pointed out by \cite{Feinstein_2022ApJ...925L...2F}.

\section{Additional plots}

\begin{figure}
\centering
\includegraphics[width=0.5\textwidth]{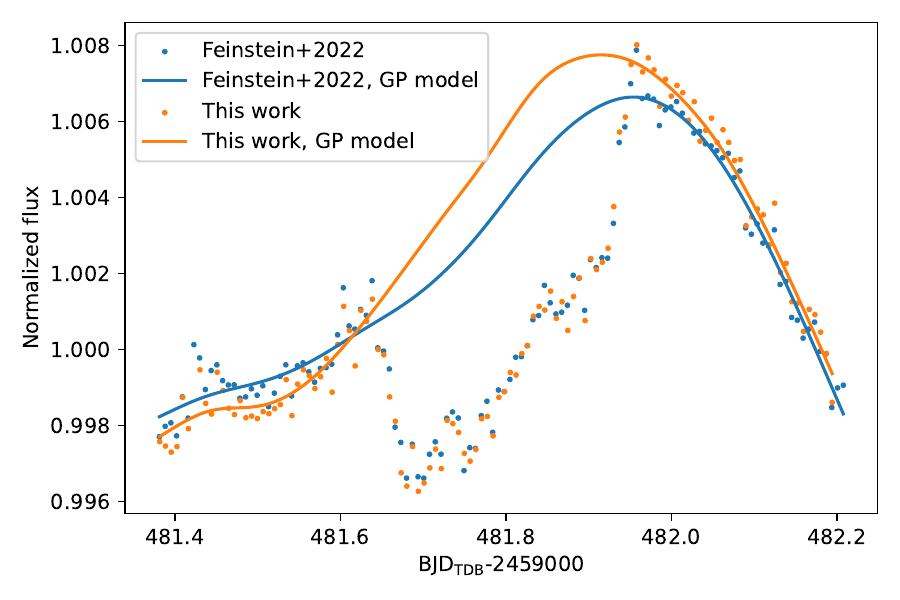}
\caption{Comparison between our extracted TESS LC (orange dots) showing the transit attributed to V1298\,Tau\,$e$, and the one extracted by \cite{Feinstein_2022ApJ...925L...2F} (blue dots), including the corresponding best-fit GP models (solid curves). }
\label{fig:tessgpcomparison}
\end{figure}

\begin{figure}
    \centering
    \includegraphics[width=0.5\textwidth]{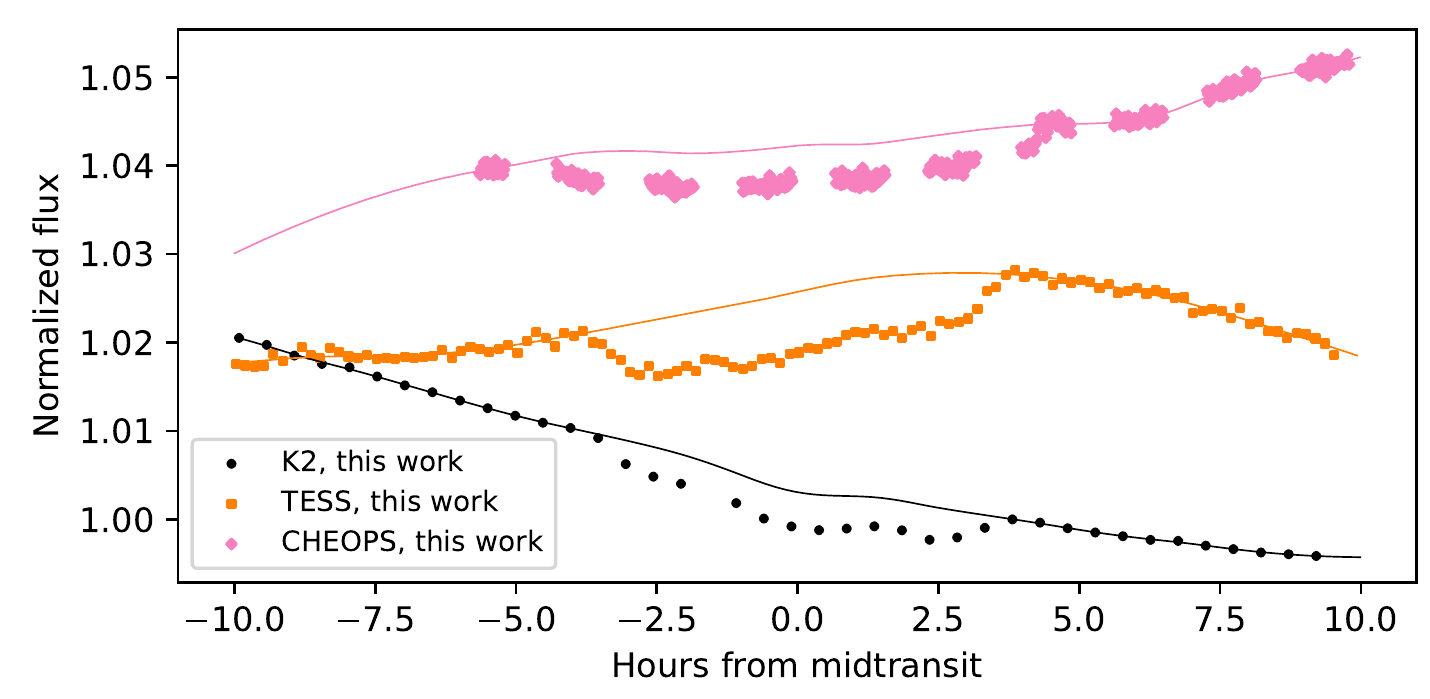}
    \caption{Undetrended LCs of the three transits analysed in our work and shown in Fig. \ref{fig:transits} ($Kepler/K2$, TESS, and CHEOPS), with overplotted our calculated best-fit GP models. An offset has been added to each curve in order to improve the clarity of the plot. }
    \label{fig:threetransitsgp}
\end{figure}

\begin{figure}
    \centering
    \includegraphics[width=0.5\textwidth]{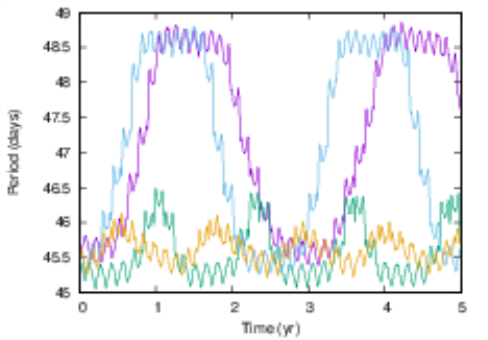}
    \caption{Temporal evolution of the orbital period of V1298\,Tau\,$e$, assuming that $K2/Kepler$, TESS, and CHEOPS observed a transit of the same planet. This illustrative result is based on a few simulated N-body integrations covering a time span of five years, as described in Section \ref{sec:conclusions}. Each curve in the plot represents a simulation.}
    \label{fig:dynam_integr}
\end{figure}

\begin{figure}
    \centering
    \includegraphics[width=0.5\textwidth]{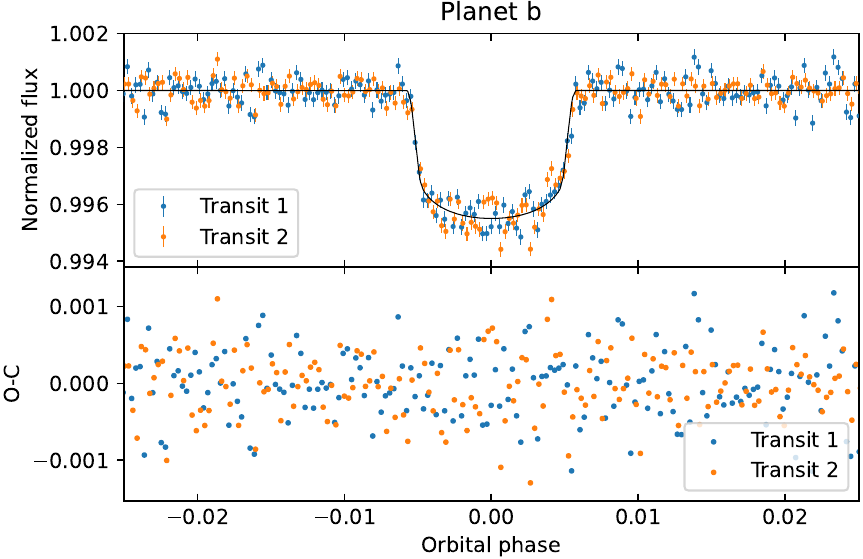}
    \caption{Comparison between the two consecutive transits of V1298\,Tau\,$b$ observed by TESS around epochs BJD$_{\rm TDB}$ 2\,459\,481 (transit 1) and 2\,459\,505 (transit 2).}    
    \label{fig:transits_b_tess}
\end{figure}

\end{appendix}

\end{document}